\documentclass[a4paper,11pt]{article}%
\usepackage{amsmath}
\usepackage{graphicx}
\usepackage{amsfonts}
\usepackage{amssymb}%
\newtheorem{theorem}{Theorem}

\newtheorem{definition}[theorem]{Definition}

\newtheorem{proposition}[theorem]{Proposition}

\newtheorem{remark}[theorem]{Remark}

\begin{document}

\title{A Unified Scheme for Generalized Sectors \\based on Selection Criteria \\--Order parameters of symmetries and of thermality and physical meanings of adjunctions--}
\author{Dedicated to the memory of Mosh\'e Flato\\
\\
\\Izumi Ojima\\Research Institute for Mathematical Sciences, \\Kyoto University, Kyoto 606-8502, Japan}
\date{}
\maketitle

\begin{abstract}
A unified scheme for treating generalized superselection sectors is proposed
on the basis of the notion of selection criteria to characterize states of
relevance to each specific domain in quantum physics, ranging from the
relativistic quantum fields in the vacuum situations with unbroken and
spontaneously broken internal symmetries, through equilibrium and
non-equilibrium states to some basic aspects in measurement processes. This is
achieved by the help of \textit{c} $\rightarrow$ \textit{q} and \textit{q}
$\rightarrow$ \textit{c} channels: the former determines the states to be
selected and to be parametrized by the order parameters, and the latter
provides the physical interpretations of selected states in terms of order
parameters. This formulation extends the traditional range of applicability of
the Doplicher-Roberts construction method for recovering the field algebra and
the gauge group (of the first kind) from the data of group invariant
observables to the situations with spontaneous symmetry breakdown: in use of
the machinery proposed, the physical and mathematical meaning of basic
structural ingredients associated with the spontaneously broken symmetry are
re-examined, such as the degenerate vacua parametrized by the variable
belonging to the relevant homogeneous space, the Goldstone modes and
condensates, etc. The geometrical meaning of the space of order parameters is
naturally understood in relation with the adjunction as the classifying space
of a sector structure. As further examples of applications, some basic notions
arising in the mathematical framework of quantum theory are reformulated and
examined in connection with control theory. \newline\newline%

\end{abstract}

\section{Introduction}

The standard way of treating the microscopic world on the basis of quantum
field theory (QFT for short) is to introduce first the \textit{quantum fields}
whose characterization is given by means of their behaviours under the various
kinds of \textit{symmetries}; e.g., the internal symmetry groups such as
colour $SU(3)$, chiral $SU(2)$, electromagnetic $U(1)$, or any other bigger
(super)groups of grand unifications (and their corresponding versions of local
gauge symmetries), in combination with spacetime symmetry groups such as
Poincar\'{e} group in Minkowski spacetime, conformal groups in massless
theories, or isometry groups of curved spacetimes, and so on. In a word, basic
objects of such a system can essentially be found in an algebra $\mathfrak{F}$
of quantum fields (called a \textit{field algebra} for short) acted upon by
two kinds of symmetries, \textit{internal }and \textit{spacetime} (whose
unification has been pursued as one of the ultimate goals of microscopic
physics). With respect to the group of an internal symmetry denoted
generically by $G$, the generating elements of $\mathfrak{F}$ (usually called
\textit{basic} or \textit{fundamental fields}) are assumed (\textit{by hand})
to belong to certain multiplet(s) transforming covariantly under the action of
$G$, which defines mathematically an action $\tau$ of $G$ on $\mathfrak{F}$:
$G\underset{\tau}{\curvearrowright}\mathfrak{F}$.

Contrary to this kind of \textit{theoretical }setting, what can be observed
(experimentally) in the real world is believed (or, can be proved under a
certain setup; for instance, see \cite{Oji78}) to be only elements in
$\mathfrak{F}$ \textit{invariant under }$G$ and are usually called
\textit{observables} which constitute the algebra $\mathfrak{A}$ of
observables:
\begin{equation}
\mathfrak{A}:=\mathfrak{F}^{G}=\{A\in\mathfrak{F};\tau_{g}(A)=A\text{ for
}\forall g\in G\},
\end{equation}
the fixed-point subalgebra of $\mathfrak{F}$ under the action $\tau$ of $G$.
Thus, what we can directly check experimentally is supposed to be only those
data described in terms of $\mathfrak{A}$ (and its derived objects) and the
rest of the notions appearing in our framework are just mathematical devices
whose pertinence can be justified only through the information related to
$\mathfrak{A}$. Except for the systematic approaches \cite{DHR, DR90, Haag}
undertaken by the pioneers in algebraic QFT \cite{Haag}, however, there have
so far been no serious attempts to understand the basic mechanism pertaining
to this point as to how a particular choice of $\mathfrak{F}$ and $G$ can be
verified, with the problems of this sort left aside just to the heuristic
arguments based on trials and errors. While a particularly chosen combination
$G\underset{\tau}{\curvearrowright}\mathfrak{F}$ is no doubt meaningless
without good agreements of its consequences with the observed data described
in terms of $\mathfrak{A}$, the attained agreements support the postulated
theoretical assumption only as \textit{one of many possible candidates} of
explanations, without justifying it as a \textit{unique} inevitable solution.
(Does it not look quite strange that such a kind of problems as this have
hardly been examined in the very sophisticated discussions about the unicity
of the unification models at the Planck scale?)

Just when restricted to the cases with $G$ an \textit{unbroken} global gauge
symmetry (or, gauge symmetry of the first kind), a satisfactory framework in
this context has been established in the \textit{superselection theory} of
Doplicher-Haag-Roberts (DHR) \cite{DHR} and of Doplicher-Roberts (DR)
\cite{DR89, DR90} (DHR-DR sector theory for short) in algebraic QFT, whose
physical essence has, unfortunately, not been recognized widely (which may be
partly due to its mathematical sophistication, but mainly due to the lack of
common understanding of the importance of the above-mentioned problem). What
is marvelous about this theory is that it enables one to \textit{recover both}
$\mathfrak{F}$ \textit{and} $G$ starting only from the data encoded in
$\mathfrak{A}$ when supplemented by the so-called DHR \textit{selection
criterion} \cite{Borch, DHR} to choose physically relevant states with
localizable charges (which need, in the case of topological charges, be
modified as in \cite{BF82}). Then the vacuum representation of the so
constructed field algebra $\mathfrak{F}$ is decomposed into mutually disjoint
irreducible representations of $\mathfrak{A}=\mathfrak{F}^{G}$, called
\textit{superselection sectors} (or \textit{sectors} for short), in one-to-one
correspondence with mutually disjoint irreducible unitary representations of
the internal symmetry group $G$ which is found to be \textit{compact Lie}.
While the essence of this theory will be briefly summarized in Sec.3 in a
reformulated form convenient for the present context, it may be meaningful to
mention some general aspects of it for the sake of explaining the reason why
we think the analysis of spontaneous symmetry breakdown (SSB for short) as important.

Among the most important consequences of the DHR-DR sector theory, we mention
here that the familiar Bose/Fermi statistics of the basic fields is
automatically derived from the \textit{local net structure $\mathcal{O}$%
}$\longmapsto\mathfrak{A}(\mathcal{O})$ consisting of\textit{ local
subalgebras }$\mathfrak{A}(\mathcal{O})$ \textit{of observables} in spacetime
regions $\mathcal{O}$ satisfying the\textit{\ local commutativity} (i.e.,
Einstein causality), without necessity of introducing from the outset
\textit{unobservable} field operators such as \textit{fermionic fields}
subject to local \textit{anti}commutativity \textit{violating Einstein
causality}; this shows that fermionic fields are, in a sense, simple
mathematical devices for bookkeeping of half-integer spin states. Then all the
non-trivial spacetime behaviours are described here by the observable net
\textit{$\mathcal{O}$}$\longmapsto\mathfrak{A}(\mathcal{O})$, while the
internal symmetry aspects are encoded in the \textit{sector structure}, which
also originates from the observable net.

The symmetry arising from this beautiful theory is, however, destined to be
\textit{unbroken},\textit{\ }excluding the situation of SSB, which poses a
question about the \textquotedblleft\textbf{stability}\textquotedblright\ of
this method, as remarked by the late Mosh\'{e} Flato \cite{Flato96}. Indeed we
know that many (actually, almost all) of the \textquotedblleft sacred
symmetries\textquotedblright\ in nature can be broken (explicitly or
spontaneously) in various situations: e.g., SSB's of chiral symmetry in the
electro-weak theory based upon $SU(2)\times U(1)$, electromagnetic $U(1)$ in
the superconductivity, and the rotation symmetry $SO(3)$ in ferromagnetism,
etc. \noindent So, the question as to whether or not this theory can
incorporate systematically the cases of SSB is a real challenge to it,
deserving serious examination, and if the answer is yes, what kind of sector
structure is realized in that case is another non-trivial interesting
question. This sort of investigation is expected also to give us some
important hints for getting rid of another restriction of \textit{global
}gauge symmetries so as to incorporate \textit{local} gauge symmetries.

In the following, we give an affirmative answer to the above question,
clarifying the sector structure emerging from SSB. For this purpose, we note
that the traditional notion of sector structures has hinged strongly to the
essential features of unbroken symmetry, allowing only the \textit{discrete
sectors }which are parametrized by the discrete $\hat{G}$, the dual of a
compact group defined as the set of all equivalence classes of
finite-dimensional continuous unitary irreducible representations of $G$. When
we start to extend this formalism to the situations with SSB, we encounter the
presence of \textit{continuous sectors} (or, \textquotedblleft degenerate
vacua\textquotedblright\ in the traditional terminology) parametrized by
continuous \textit{macroscopic order parameters}, as is seen in Sec.4.

This forces us to extend the notion of sectors to incorporate the continuous
ones. Once we take this viewpoint, then we notice that unexpectedly wide
perspectives open out before us: aside from such very fundamental issues as
the \textquotedblleft ultimate\textquotedblright\ unifications, we have so far
faced with so many different levels and domains of physical nature in the
directions from microscopic worlds to macroscopic ones, ranging from the
vacuum situations (the standard QFT relevant to particle physics), thermal
equilibria (QFT at finite temperatures or quantum statistical mechanics),
non-equilibrium ones and so on, where we find a huge supply of examples of
continuous sectors. In \cite{BOR01}, a general framework is proposed for
defining non-equilibrium local states in relativistic QFT and for describing
their thermodynamic properties in terms of the associated macroscopic
observables found in the \textit{centre} of a (kind of) \textquotedblleft
universal\textquotedblright\ representation containing all the representations
of observables relevant to the context. From this general standpoint, one
easily notices that the thermal equilibria at different temperatures can be
seen to constitute families of continuous sectors parametrized by such
thermodynamic variables as temperatures, chemical potentials and pressure and
so on. In view of such roles of central observables associated with continuous
sectors appearing in SSB cases as well as the above various kinds of thermal
states, it is appropriate now to try the possibility of unified ways of
treating these different cases, just regarding the traditional discrete ones
as special cases; this is simply parallel to the extension of the traditional
\textit{eigenvalue} problems for linear operators with \textit{discrete
spectra }to the general \textit{spectral decompositions }admitting the
appearance of \textit{continuous spectra}.

Thus the aim of the present paper is threefold: to incorporate SSB into the
sector theory, we extend the notion of discrete sectors to continuous ones,
through which we are led to a unified scheme to treat such a generalized
notion of sectors. The key roles are played by the \textit{selection criterion
}set up at the starting point of theory in defining and choosing
\textit{physically relevant family of states} as well as in providing a
systematic way for \textit{describing} and \textit{interpreting }relevant
physical properties.

We introduce the necessary ingredients for formulating the scheme through the
discussions on the basic structures found in thermal situations of equilibrium
and of the extension to non-equilibrium (Sec.2) and in an operational
reformulation of DHR-DR sector theory (Sec.3). Here the general mathematical
meaning of selection criterion is found in the \textit{adjunction} as a tool
for controlling the mutual relations between generic objects to be
characterized and special objects (such as temperatures or order parameters,
in general) serving as standard reference systems in the context of
classification and interpretation.

In Sec.4, we apply our unified method of treating generalized sectors to the
situation with SSB by combining the discussion in Sec.2 for continuous sectors
and one in Sec.3 for discrete ones. What is interesting physically and
mathematically is the duality relation (states $\longleftrightarrow$ algebra)
between the above degenerate vacua (as \textit{states}) with global classical
parameters and the local quantum Goldstone modes found in the dual-net
\textit{algebra} $\mathfrak{A}^{d}$ of extended observables in a vacuum
representation, in view of its close relation to such a physical picture of
Goldstone modes to search degenerate vacua in a virtual way (i.e., while the
order parameter $G/H$ of SSB from $G$ down to $H$ is exhibited as a
macroscopic quantity by the degenerate vacua, the quantum Golstone modes
$\varphi$ related to $G/H$ represent in a fixed irreducible representation the
virtual transitions from a vacuum to another, as seen in the analysis of the
Goldstone commutators, $\omega(\delta_{X}(\varphi))\neq0$, as the
infinitesimal form of symmetry breaking $\omega(\tau_{g}(\varphi))\neq
\omega(\varphi)$ for $g\in G\diagdown H$). At the same time, this is also
related with the mathematical notion of duality between a homogeneous space
$G/H$ and its representations, as a natural extension of Tannaka-Krein duality
of compact groups \cite{TK}. Then the basic structural features of the theory
with spontaneously broken symmetry are clarified, establishing mutual
relationship among degenerate vacua, order parameters, Goldstone modes and
condensates responsible for SSB (see Sec.4.3 and 4.4). Since these constitute 
the starting points for the systematic approach, there are many things to be settled and developed further as is indicated. 

In Sec.5, we explain the general mathematical meaning of the proposed scheme,
in relation with the categorical adjunctions, especially with the geometric
notions of classifying spaces and classifying maps. As further examples of
applications of the method, we examine also some basic notions supporting the
physical and operational meanings of the mathematical framework of quantum
theory; as the standard probabilistic interpretation is naturally understood
as something arising from physical processes between measured objects and
measuring systems, we can formulate and examine the measurability problem of
particular physical quantities as the \textit{realizability }of certain
physical dynamical processes suited for the purpose, which is among the
typical problems appearing in the context of control theory. In the same
context, the problem of state preparation can be treated as a
\textit{reachability} problem to examine whether there is a process to bring
the system into any desired condition. In both cases, we find that what to be
selected is not always states but can also be channels.

\section{Selection criteria and \textit{c}$\rightarrow$\textit{q} \&
\textit{q}$\rightarrow$\textit{c} channels in thermal situations}

\subsection{Equilibrium states and thermal interpretations}

To draw a clear picture of the idea, we briefly sketch the essense of the
scheme proposed in \cite{BOR01} for defining and describing non-equilibrium
local states in a relativistic QFT. From the present standpoint, it can be
reformulated as follows according to \cite{Oji02, Oji03}. To characterize an
unknown state $\omega$ as a non-equilibrium local state, we prepare the
following basic ingredients.

\begin{itemize}
\item[i)] Candidates of such states are sought within the set $E$ of states
$\omega$ (understood as an \textit{expectation functional},\textit{
}mathematically formulated as a normalized positive linear functional on the
algebra $\mathfrak{A}$ of observables of the system under consideration) with
locally finite energy characterized by the energy-bound condition \cite{FrHe}
\begin{equation}
\omega((\mathbf{1}+H_{\mathcal{O}})^{2m})<\infty\label{regularity}%
\end{equation}
valid in some spacetime local region $\mathcal{O}$ and some $m>0$ with
$H_{\mathcal{O}}$ a \textit{local Hamiltonian} playing the role of Hamiltonian
in $\mathcal{O}$ (whose definition is justified under the assumption of the
nuclearity condition \cite{BDL}). This choice is so designed that the
comparison is fully meaningful between an unknown state $\omega\in E$ and
known reference states $\in K$ specified in the next ii) as statistical
mixtures of thermal equilibria, \textit{in infinitesimally small
neighbourhoods of a spacetime point} $x$ by means of observables
$\in\mathcal{T}_{x}$ defined in iii).) We denote $E_{\mathcal{O}}$ the totally
of states $\omega$ satisfying Eq.(\ref{regularity}) with a suitable $m>0$,
\begin{equation}
E_{\mathcal{O}}:=\{\omega;\omega\text{: state of }\mathcal{A}\text{ and
}\exists m>0\text{ s.t. }\omega((\mathbf{1}+H_{\mathcal{O}})^{2m})<\infty\},
\end{equation}
whose pointlike limit (projective limit)
\begin{equation}
E_{x}(=\underset{\mathcal{O}\rightarrow x}{\underleftarrow{\lim}%
}E_{\mathcal{O}}) \label{germ}%
\end{equation}
is given by the set of equivalence classes in $\cup_{\mathcal{O}%
}E_{\mathcal{O}}$ with respect to the equivalence relation $\thicksim
$\ defined by
\begin{equation}
\omega_{1}\thicksim\omega_{2}\overset{\mathrm{def}}{\Longleftrightarrow
}\exists\mathcal{O}\text{: neighbourhood of }x\text{ s.t. }\omega
_{1}\upharpoonright_{\mathcal{O}}=\omega_{2}\upharpoonright_{\mathcal{O}}.
\end{equation}

\begin{proposition}
If the local Hamiltonians $H_{\mathcal{O}}$ are positive, the family
$\mathcal{O}\longmapsto E_{\mathcal{O}}$ constitutes a presheaf of state germs
\cite{HO} whose stalk at $x$ is given by $E_{x}$.
\end{proposition}

(Proof is simple and omitted.)

\item[ii)] The set $K$ of \textit{thermal reference states }consisting of all
global thermal equilibrium states defined as the relativistic KMS states
$\omega_{\beta}$ \cite{BrBu} (with inverse temperature 4-vectors $\beta
=(\beta^{\mu})\in V_{+}:=\{x\in\mathbb{R}^{4};x^{0}>0,x^{2}=(x^{0})^{2}%
-\vec{x}^{2}>0\}$) and of their suitable convex combinations: $K$ plays the
role of a \textit{model space }whose analogue in the definition of a manifold
$M$ can be\textit{\ }found in a Euclidean space $\mathbb{R}^{n}$ as the value
space of local charts. Any states belonging to this set $K$ is seen to belong
to the above $E_{\mathcal{O}}$ with any arbitrary finite spacetime region
$\mathcal{O}$: $K\subset E_{\mathcal{O}}\subset E$.

\item[iii)] The linear space $\mathcal{T}_{x}$ of \textit{local thermal
observables}\footnote{While the set $\mathcal{T}_{x}$ is designed for
detecting local thermal properties in use of quantum observables, it is
defined as a suitable subset of the point-like fields dual to the state germs
\cite{HO, Bo}, relying essentially on the criterion \cite{Borch}\ to select
states with moderate energy contents. The elements of \textit{thermality
}enters here only in excluding certain point-like observables irrelevant to
thermal contexts, and hence, the naming with \textquotedblleft\textit{thermal}%
\textquotedblright\textit{ }may somehow be misleading, as remarked by Prof. R.
Haag.} is defined as linear forms on states in $E_{x}$ satisfying the
regularity (\ref{regularity}) which makes meaningful the notion of
\textit{quantum fields at a point }$x$ \cite{BOR01, Bo}:
\begin{equation}
\mathcal{T}_{x}\,:=\,\sum_{p,q}\,\mathcal{N}(\hat{\phi}_{0}^{\,p})_{\,q,x}\,,
\end{equation}
where $\hat{\phi}_{0}^{\,}$ generically denotes the basic quantum fields
defining our QFT. The notion of \textit{normal products }$\mathcal{N}%
(\hat{\phi}_{0}^{\,p})_{\,q,x}$ enters here to recover effectively the product
structure of quantum fields lost through the process of pointlike limit,
arising from the operator product expansion (OPE) of $\hat{\phi}_{0}%
^{\,}(x+\zeta_{1})\cdots\hat{\phi}_{0}^{\,}(x+\zeta_{p})$ in the limit of
$\zeta_{i}\rightarrow0$, $\sum_{j}\zeta_{j}=0$ reformulated recently by
\cite{Bo} in a mathematically rigorous form. The simplest case, $\mathcal{N}%
(\hat{\phi}^{2})_{\,q,x}$, with $p=2$ can be understood as the linear space
spanned by the coefficients $\hat{\Phi}_{j}(x)$ of $c$-number singular
functions $c_{j}(\zeta)$ in $\zeta$ in%
\begin{equation}
||(\mathbf{1}+H_{\mathcal{O}})^{-n}\left[  \hat{\phi}(x+\zeta)\hat{\phi
}(x-\zeta)-\sum_{j=1}^{J(q)}c_{j}(\zeta)\,\hat{\Phi}_{j}(x)\right]
(\mathbf{1}+H_{\mathcal{O}})^{-n}||\leq c^{\prime}\,|\zeta|^{q},
\label{wilson}%
\end{equation}
valid for sufficiently large $n\in\mathbb{N}$, which serve as substitutes for
the ill-defined $\hat{\phi}(x)^{2}$, and similarly $\mathcal{N}(\hat{\phi}%
^{p})_{\,q,x}$ for higher power $\hat{\phi}(x)^{p}$. What is important about
$\mathcal{T}_{x}$ is its natural \textit{hierarchical structure} ordered by
the indices $p,q$ related to energy bound and OPE, starting from scalar
multiples of identity to higher powers $\mathcal{N}(\hat{\phi}^{p})_{\,q,x}$
with the larger $p$ providing the finer resolution. \newline Along the above
analogy to a manifold $M$ in differential geometry, their role is to relate
our unknown state $\omega\in E$ to the known reference states in $K$, just in
parallel to the local coordinates which relate locally a generic curved space
$M$ to the known space $\mathbb{R}^{n}$. As explained just below, the physical
interpretations of local thermal observables $\hat{A}$ are given by
macroscopic \textit{thermal functions} $A$ corresponding to $\hat{A}%
$,\footnote{Whenever convenient without fear of confusions, we adopt here a
physicist's convention to indicate the correspondence and distinction between
a quantum observable $\hat{A}$ and a classical one $A$ in a suitable
correspondence to the former.} through which our unknown $\omega$ can be
\textit{compared} with thermal reference states in $K$.
\end{itemize}

Before going into the discussion of non-equilibrium, we need first to
establish the physical roles of the above ingredients for describing the
relevant thermal properties of states and quantum observables in the realm $K$
of generalized thermal equilibria. To this end, we introduce

\begin{definition}
\textit{\textbf{Thermal functions }}are defined for each quantum observables
$\hat{A}$($\in\mathcal{T}_{x}$) by the map
\begin{align}
\mathcal{C}  &  :\hat{A}\longmapsto\mathcal{C}(\hat{A})\in C(B_{K})\text{
\ }\nonumber\\
&  \text{with\ }\mathcal{C}(\hat{A})(\beta,\mu):=\omega_{\beta,\mu}(\hat
{A})\text{ \ for\ \ }(\beta,\mu)\in B_{K},
\end{align}
where $B_{K}$ is the classifying space to parameterize thermodynamic pure
phases, consisting of inverse temperature 4-vectors $\beta\in V_{+}$ in
addition to any other thermodynamic parameters (if any) generically denoted by
$\mu$ (e.g., chemical potentials) necessary to exhaust and discriminate all
the thermodynamic pure phases.
\end{definition}

Since the map $\mathcal{C}$ is easily seen to be unital and positive linear,
$\mathcal{C}(\mathbf{1})=1,\mathcal{C}(\hat{A}^{\ast}\hat{A})\geq0$%
,\thinspace\ it is a \textit{completely positive} map characterized by the
condition $\sum_{ij=1}^{n}\bar{f}_{i}\mathcal{C}(\hat{A}_{i}^{\ast}\hat{A}%
_{j})f_{j}\geq0$ for $\forall n\in\mathbb{N},\forall f_{1},\cdots,\forall
f_{n}\in C(B_{K})$ (and $\forall\hat{A}_{i}$'s belonging to a suitable
C*-algebra $\mathfrak{A}$ to which the operator space $\mathcal{T}_{x}$ is
affiliated). As the dual of a completely positive map, $\mathcal{C}^{\ast}$ on
states becomes a \textit{\textbf{c}lassical-\textbf{q}uantum (c}$\rightarrow
$\textit{q) channel} \cite{OhyaPetz} $\mathcal{C}^{\ast}:Th\ni\rho
\longmapsto\mathcal{C}^{\ast}(\rho)\in K$ given by%
\begin{align}
&  \mathcal{C}^{\ast}(\rho)(\hat{A})=\rho(\mathcal{C}(\hat{A}))=\int_{B_{K}%
}d\rho(\beta,\mu)\mathcal{C}(\hat{A})(\beta,\mu)=\int_{B_{K}}d\rho(\beta
,\mu)\omega_{\beta,\mu}(\hat{A}),\nonumber\\
&  \Longrightarrow\mathcal{C}^{\ast}(\rho):=\int_{B_{K}}d\rho(\beta,\mu
)\omega_{\beta,\mu}=\omega_{\rho}\in K. \label{Decomp}%
\end{align}
Here $Th:=M_{1}(B_{K})$ is the space of classical thermal states identified
with probability measures $\rho$ on $B_{K}$ describing the mean values of
thermodynamic parameters $(\beta,\mu)$ together with their fluctuations. One
can see that thermal interpretation of local quantum thermal observables
$\hat{A}\in\mathcal{T}_{x}$ is given in all thermal reference states of the
form $\mathcal{C}^{\ast}(\rho)=\omega_{\rho}\in K$ by the corresponding
thermal function $\mathcal{C}(\hat{A})$ evaluated with the classical
probability $\rho$ describing the thermodynamic configurations of
$\omega_{\rho}$ through the relation
\begin{equation}
\omega_{\rho}(\hat{A})=\int_{B_{K}}\!d\rho(\beta,\mu)\,\omega_{\beta,\mu}%
(\hat{A})=\rho(\mathcal{C}(\hat{A})).
\end{equation}
This applies to the case where $\rho$ is already known. What we need to ask in
the actual situations is how to determine the unknown $\rho$ from the given
data set $\Phi\longmapsto\rho(\Phi)$ of expectation values of thermal
functions $\Phi$ (which is the problem of state estimation): this problem can
be solved if $\mathcal{T}_{x}$ has sufficiently many local thermal observables
so that the totality $\mathcal{C}(\mathcal{T}_{x})$ of the corresponding
thermal functions can approximate arbitrary continuous functions of
$(\beta,\mu)\in B_{K}$. In this case $\rho$ is given as the unique solution to
a (generalized) \textquotedblleft moment problem\textquotedblright. Thus we see:

\begin{itemize}
\item[$\bigstar$] If the set $\mathcal{T}_{x}$\textit{\ }of local thermal
observables is large enough to discriminate\ all the thermal reference states
in $K$, then any reference state $\in K$ can be written as $\mathcal{C}^{\ast
}(\rho)$ in terms of a uniquely determined probability measure $\rho$ on
$B_{K}$ describing the statistical fluctuations of thermal parameters in the
state in question. Then local thermal observables\textit{\ }$\hat{\Phi}%
\in\mathcal{T}_{x}$\textit{\ }provide the same information on the thermal
properties of states in $K$ as that provided by the corresponding classical
macroscopic thermal functions\textit{\ }$\Phi=\mathcal{C}(\hat{\Phi})$\ [e.g.,
internal energy, entropy density, etc.]: $\omega_{\rho}(\hat{\Phi})=\rho
(\Phi)$.
\end{itemize}

In this situation, \textit{any continuous function} $F$ in $B_{K}$ can be
approximated by thermal functions $\Phi_{x}=\mathcal{C}(\hat{\Phi}(x))$ with
arbitrary precision, \textit{even if} $F$ itself is \textit{not} an image of
$\mathcal{C}$:
\begin{equation}
\overline{\mathcal{C}(\mathcal{T}_{x})}^{||\cdot||}=C(B_{K}). \label{dense}%
\end{equation}
For instance, the \textit{entropy density} $s(\beta)$ can be treated as such
an \textit{approximate} thermal function in spite of the absence of quantum
observables $\hat{s}(x)\in\mathcal{T}_{x}$ s.t. $\omega_{\beta}(\hat
{s}(x))=s(\beta)$. What the above ($\bigstar$) says is the equality and the
equivalence,
\begin{align}
K  &  =\mathcal{C}^{\ast}(Th);\\
\omega_{\rho_{1}}\underset{\mathcal{T}_{x}}{\equiv}\omega_{\rho_{2}}  &
\Longleftrightarrow\rho_{1}\underset{\mathcal{C}(\mathcal{T}_{x})}{\equiv}%
\rho_{2}, \label{equiv}%
\end{align}
for $\rho_{i}\in Th$, $\omega_{\rho_{i}}=\mathcal{C}^{\ast}(\rho_{i}%
)=\int_{B_{K}}d\rho_{i}(\beta,\mu)\omega_{\beta,\mu}\in K$, where
$\underset{\mathcal{T}_{x}}{\equiv}$ and $\underset{\mathcal{C}(\mathcal{T}%
_{x})}{\equiv}$ denote the equivalence relations in $K$ and $Th$ given
respectively by
\begin{align}
\omega_{1}\underset{\mathcal{T}_{x}}{\equiv}\omega_{2}  &  \Longleftrightarrow
(\omega_{1}-\omega_{2})(\mathcal{T}_{x})=\{0\},\label{equiv1}\\
\rho_{1}\underset{\mathcal{C}(\mathcal{T}_{x})}{\equiv}\rho_{2}  &
\Longleftrightarrow(\rho_{1}-\rho_{2})(\mathcal{C}(\mathcal{T}_{x}))=\{0\}.
\label{equiv2}%
\end{align}
So, it ensures the existence of \textit{inverse of c}$\rightarrow$\textit{q
channel} $\mathcal{C}^{\ast}$ on $K$:
\begin{equation}
K\ni\omega_{\rho}=\mathcal{C}^{\ast}(\rho)\longleftrightarrow(\mathcal{C}%
^{\ast})^{-1}(\omega_{\rho})=\rho\in Th, \label{Fourier}%
\end{equation}
and the thermal interpretation of thermal reference states $\in K$ is just
given by this \textbf{\textit{\textbf{q} }}$\rightarrow$\textit{\textbf{c
channel }}$(\mathcal{C}^{\ast})^{-1}:K\ni\omega\longmapsto\rho\in Th$ s.t.
$\omega=\mathcal{C}^{\ast}(\rho)$ \cite{Oji02}. In the parallelism between the
integral representation in Eq.(\ref{Decomp}) and the Fourier decomposition of
a function, we note that $(\mathcal{C}^{\ast})^{-1}$ acting on $\omega_{\rho
}\in K$ corresponds to the Fourier transform.

To adapt to our discussion of local thermal situations, we summarize the above
points in such a form of \textit{adjunction }\cite{MacL}\textit{ }as
\begin{equation}
K/\mathcal{T}_{x}(\omega,\mathcal{C}^{\ast}(\rho))\overset{q\rightleftarrows
c}{\simeq}Th/\mathcal{C}(\mathcal{T}_{x})((\mathcal{C}^{\ast})^{-1}%
(\omega),\rho), \label{adjunction1}%
\end{equation}
with a quantum state $\omega\in E$ and a probability measure $\rho\in Th$.
Since the adjunction turns out to be a convenient tool in formulating a scheme
for attaining simultaneously the selection of relevant objects (on the left)
and interpreting the selected objects (on the right), we make a slight detour
for explaining it here. While its most general formulation should be given in
the context of categories and functors (see Sec.5), we concentrate here on our
present context of treating equivalence relations given by Eqs.(\ref{equiv1})
and (\ref{equiv2}), according to which the sets $K$ and $Th$ become groupoids.

Roughly speaking, a groupoid $\Gamma$ is such a generalization of a group that
there are many unit elements which constitute a set $\Gamma_{0}$ and that the
product $\gamma_{1}\gamma_{2}$ of two elements $\gamma_{1},\gamma_{2}$ is
defined only conditionally in the following sense: Each element (also called
an \textquotedblleft arrow\textquotedblright) $\gamma\in\Gamma$ has its source
$s(\gamma)$ and target $r(\gamma)$ in $\Gamma_{0}$ and these points are
thought to be connected by $\gamma$, $s(\gamma)\overset{\gamma}{\rightarrow
}r(\gamma)$ (dented also as $\gamma:s(\gamma)\rightarrow r(\gamma)$), in an
invertible way: $r(\gamma)\overset{\gamma^{-1}}{\rightarrow}s(\gamma)$. Two
arrows $\gamma_{1},\gamma_{2}\in\Gamma$ are composable to yield a product
$\gamma_{1}\gamma_{2}\in\Gamma$ if and only if $r(\gamma_{2})=s(\gamma_{1})$
and the product is to be associative: $(\gamma_{1}\gamma_{2})\gamma_{3}%
=\gamma_{1}(\gamma_{2}\gamma_{3})$. There is a one-to-one and onto
correspondence between an \textit{equivalence relation} $\thicksim$\ on a set
$\Gamma_{0}$ and a groupoid $\Gamma$ through [$a\thicksim b$ for $a,b\in$
$\Gamma_{0}$] $\Longleftrightarrow$ [$\exists\gamma\in\Gamma$ s.t.
$a=s(\gamma)$ and $b=r(\gamma)$], according to which the characterization of
equivalence relation [$a\thicksim a$], [$a\thicksim b\Longrightarrow
b\thicksim a$], [$a\thicksim b,b\thicksim c\Longrightarrow a\thicksim c$] is
translated into the basic properties of $\Gamma$ as the presence of unit
$(\overset{\iota_{a}}{a\rightarrow a})\in\Gamma$ (for $\forall a\in$
$\Gamma_{0}$), the invertibility of any $\gamma$: $\Gamma\ni(\overset{\gamma
}{a\rightarrow b})\Longrightarrow(\overset{\gamma^{-1}}{b\rightarrow a}%
)\in\Gamma$, and the composition: $(\overset{\gamma_{1}}{a\rightarrow
b}),(\overset{\gamma_{2}}{b\rightarrow c})\in\Gamma\Longrightarrow
(\overset{\gamma_{2}\gamma_{1}}{a\rightarrow c})\in\Gamma$. Collecting all the
arrows from $a\in$ $\Gamma_{0}$ to $b\in$ $\Gamma_{0}$, we denote
$\Gamma(a,b):=\{\gamma\in\Gamma;s(\gamma)=a$ and $r(\gamma)=b\}$. Viewed as a
category, $\Gamma$ is one with $\Gamma_{0}$ as the set of objects and with all
its arrows being invertible. Corresponding to the equivalence relations
$\underset{\mathcal{T}_{x}}{\equiv}$ and $\underset{\mathcal{C}(\mathcal{T}%
_{x})}{\equiv}$ defined by Eqs. (\ref{equiv1}) and (\ref{equiv2}) on $K$ and
$Th$, we can consider the groupoids denoted respectively by $K/\mathcal{T}%
_{x}$ and $Th/\mathcal{C}(\mathcal{T}_{x})$. Then

\begin{proposition}
Under the condition of ($\bigstar$), the groupoids $K/\mathcal{T}_{x}$ and
$Th/\mathcal{C}(\mathcal{T}_{x})$ are isomorphic with the \textit{c}%
$\rightarrow$\textit{q }channel $\mathcal{C}^{\ast}:Th\rightarrow K$ as a
groupoid isomorphism preserving the structures as in (\ref{equiv}).
\end{proposition}

For an arbitrary state $\omega\in E_{x}$ at $x$, the existence of a non-empty
set $K/\mathcal{T}_{x}(\omega,\mathcal{C}^{\ast}(\rho))$ of arrows in
$K/\mathcal{T}_{x}$ identifies it with a uniquely determined member
$\mathcal{C}^{\ast}(\rho)$ of $K$ through the relation $\omega\underset
{\mathcal{T}_{x}}{\equiv}$ $\mathcal{C}^{\ast}(\rho)$, which can be
transmitted by the \textit{q}$\rightarrow$\textit{c }channel $(\mathcal{C}%
^{\ast})^{-1}$ meaningful on $K$ to the right-side $Th/\mathcal{C}%
(\mathcal{T}_{x})((\mathcal{C}^{\ast})^{-1}(\omega),\rho)$ of
Eq.(\ref{adjunction1}) to provide the thermal interpretation of the selected
$\omega$ by $(\mathcal{C}^{\ast})^{-1}(\omega)\underset{\mathcal{C}%
(\mathcal{T}_{x})}{\equiv}$ $\rho\in Th$ in terms of a probability
distribution $\rho$ (of temperature, etc.).

While the use of adjunction may look like something unnecessarily pedantic in
the present simple situation treating equivalence relations, the essence of
($\bigstar$) in this form (\ref{adjunction1}) can be generalized to wider
contexts as a selection criterion to choose states of relevance. Then we will
encounter more involved cases where the arrows of relevant categories are not
necessarily invertible and the \textit{c}$\rightarrow$\textit{q }and\textit{
q}$\rightarrow$\textit{c }channels need be replaced by \textit{functors}
constituting \textit{adjoint pairs} and so on. One of the merits of the use of
adjunctions is that it clearly shows the characteristic features, essence, and
basic ingredients common to all the problems to \textit{select} objects with
specific properties from generic ones and to \textit{describe, interpret} and
\textit{classify} the features of all what to be selected by comparing them
with special standard reference objects. In this setup, for instance, it is
evident and conceptually important that we have here \textit{two different
levels or domains}, quatum statistical mechanics with family $K$ of mixtures
of KMS states and macroscopic thermodynamics described by $Th$ of probability
measures of fluctuating thermal parameters on the parameter space $B_{K}$,
which are so interrelated by the two channels,\textit{\textbf{\ }c
}$\rightarrow$\textit{q} ($\mathcal{C}^{\ast}$) and \textit{q }$\rightarrow
$\textit{c }($(\mathcal{C}^{\ast})^{-1}$), that the following two points are
simultaneously attained:

a) characterization of thermal reference states $K$ as image of $\mathcal{C}%
^{\ast}$, $\omega_{\rho}=\mathcal{C}^{\ast}(\rho)$: \textit{selection
criterion} for $K$,

b) \textit{thermal interpretation }of selected states in $K$ in terms of
classical data, $\Phi_{x}=\mathcal{C}(\hat{\Phi}(x))$ and $\rho=(\mathcal{C}%
^{\ast})^{-1}(\omega_{\rho})$. \newline To implement this sort of machineries
in the actual situations, the most non-trivial steps are the pertinent
choices\textit{\textbf{ }}of pair of maps (adjoint pair of functors)
corresponding to (and, generalizing) the \textit{c}$\rightarrow$\textit{q
}and\textit{ q}$\rightarrow$\textit{c }channels\textit{\textbf{ }}together
with the \textit{\textbf{standard reference systems}} for comparison.

Going back to the original context, the problem is now boiled down into how to
select suitable classes of \textit{non-equilibrium states }$\omega\notin K$ in
such a way that some thermal interpretations are still guaranteed. This is
what to be answered in the next subsection.

\subsection{Selection criterion for non-equilibrium states}

\textit{Selection criterion and thermal interpretation of non-equilibrium
local states based on hierarchized zeroth law of local thermodynamics}\textbf{
}\cite{Oji02}: To meet simultaneously the two requirements of characterizing
an unknown state $\omega$ as a non-equilibrium local state and of establishing
its thermal interpretations in a similar way to the above a) and b), we now
compare $\omega$ with thermal reference states $\in K=\mathcal{C}^{\ast}(Th)$
by means of some local thermal observables at $x$ whose physical meanings are
exhibited by the associated thermal functions as seen above. In view of the
above conclusion [\textit{q}$\rightarrow$\textit{c} channel $(\mathcal{C}%
^{\ast})^{-1}$ on $K$] = [thermal interpretation of quantum states] and also
of the hierarchical structure in $\mathcal{T}_{x}$, we relax the requirement
for $\omega$ to agree with $\exists\omega_{\rho_{x}}:=\mathcal{C}^{\ast}%
(\rho_{x})\in K$ up to some suitable \textit{sub}space $\mathcal{S}_{x}$ of
local thermal observables $\mathcal{T}_{x}$. Then we characterize $\omega$ as
a non-equilibrium local state by

\begin{itemize}
\item[iii)] a \textit{selection criterion }for $\omega$ to be $\mathcal{S}%
_{x}$\textit{-thermal} at $x$, requiring the existence of $\rho_{x}\in Th$
s.t.
\begin{equation}
\omega(\hat{\Phi}(x))=\mathcal{C}^{\ast}(\rho_{x})(\hat{\Phi}(x))\ \text{ for
}\forall\hat{\Phi}(x)\in\mathcal{S}_{x}, \label{S-thermal}%
\end{equation}
or, $\omega\underset{\mathcal{S}_{x}}{\equiv}\mathcal{C}^{\ast}(\rho_{x})$,
for short. In terms of thermal functions $\Phi:=\mathcal{C}(\hat{\Phi}%
(x))\in\mathcal{C}(\mathcal{S}_{x})$, this can be rewritten as
\begin{equation}
\omega(\Phi)(x):=\omega(\hat{\Phi}(x))=\rho_{x}(\Phi),\quad\Phi\in
\mathcal{C}(\mathcal{S}_{x}). \label{ThermalFn2}%
\end{equation}
So, $\omega$: $\mathcal{S}_{x}$\textit{-thermal }implies that the selection
criterion $\omega\underset{\mathcal{S}_{x}}{\equiv}\mathcal{C}^{\ast}(\rho
_{x})$ can be \textquotedblleft solved\textquotedblright\ conditionally in
favour of $\rho_{x}$ as $``(\mathcal{C}^{\ast})^{-1}"(\omega)\underset
{\mathcal{C}(\mathcal{S}_{x})}{\equiv}\rho_{x}$, which provides the local
thermal interpretation of $\omega$ \cite{Oji02}. Physically this means the
state $\omega$ looks like a statistical mixture $\mathcal{C}^{\ast}(\rho_{x})$
of thermal equilibria \textit{locally} at $x$ to within a level controlled by
a subset $\mathcal{S}_{x}$ of thermal observables.
\end{itemize}

To be precise mathematically, we should be careful here about the meaning of
such a heuristic expression as $``(\mathcal{C}^{\ast})^{-1}"(\omega)$ for
$\omega\notin K$ in relation to our observation above: $\omega\notin
K=\mathcal{C}^{\ast}(Th)$. Physically this is related to the deviations of
$\omega$ from $\mathcal{C}^{\ast}(\rho_{x})$ revealed by the finer resolutions
which exhibit the extent of $\omega$ being \textit{away from equilibrium} even
locally. As we shall see below, $``(\mathcal{C}^{\ast})^{-1}"$ outside of $K$
is certainly \textit{not} a \textit{q}$\rightarrow$\textit{c} channel
preserving the positivity, whereas it can be seen to be still definable on the
states $\omega$ selected out by the above criterion Eq.(\ref{S-thermal}), by
means of its equivalent reformulation given by:

\begin{proposition}
\cite{BOR01} For a subspace $\mathcal{S}_{x}$ of $\mathcal{T}_{x}$ containing
$\mathbf{1}$, a state $\omega\in E_{x}$ is $\mathcal{S}_{x}$\textit{-thermal}
iff there is a compact set ${B}\subset V_{+}$ of inverse temperatures s.t.
\begin{align}
|\omega(\hat{\Phi}(x))|  &  \leq\tau_{B}(\hat{\Phi}(x)):=\sup_{(\beta,\mu)\in
B_{K},\beta\in B}\,|\omega_{\beta,\mu}(\hat{\Phi}(x))|\nonumber\\
&  =\left\vert \left\vert \mathcal{C}(\hat{\Phi}(x))\right\vert \right\vert
_{B},\quad\text{for }\hat{\Phi}(x)\in\mathcal{S}_{x}. \label{norm}%
\end{align}

\end{proposition}

\noindent(For the above semi-norm to be well-defined, $B_{K}\ni(\beta
,\mu)\longmapsto\omega_{\beta,\mu}\in K$ should be (weakly) continuous, which
requires singularities of critical points to be excluded from our considerations.)

Since the requirement for $``(\mathcal{C}^{\ast})^{-1}"(\omega)$ to be a
probability measure forces $\omega$ to belong to $K$, it is incompatible with
our premise $\omega\notin K$. However, the above inequality (\ref{norm})
combined with the Hahn-Banach extension theorem (under the assumption for
$\tau_{B}$ to be a norm) allows us to extend $\mathcal{C}(\mathcal{S}_{x}%
)\ni\mathcal{C}(\hat{\Phi}(x))\longmapsto\omega(\hat{\Phi}(x))$ as a
\textit{linear functional} defined on $\mathcal{C}(\mathcal{S}_{x})$ to one
$\nu$ defined on $\overline{\mathcal{C}(\mathcal{T}_{x})}=C(B_{K})$, which
should \textit{not} be a positive-definite measure but is allowed to be a
\textit{signed }measure: $\nu=\nu_{+}-\nu_{-}$, $0\leq\nu_{\pm}\in
C(B_{K})_{+}^{\ast}$, $\nu_{-}\neq0$, $\nu_{-}\upharpoonright_{\mathcal{C}%
(\mathcal{S}_{x})}=0$, $\mathcal{C}^{\ast}(\nu_{+})\upharpoonright
_{\mathcal{S}_{x}}=\omega$ $\upharpoonright_{\mathcal{S}_{x}}$. (See the
similar argument in \cite{BOR01} for the existence of an $\mathcal{S}_{x}%
$-thermal state $\omega$ showing deviations from $K$ for observables outside
of a finite-dimensional $\mathcal{S}_{x}$ as well as the treatment of the case
with $\tau_{B}$ being a \textit{semi-}norm.) Thus, understanding the meaning
of $(\mathcal{C}^{\ast})^{-1}(\omega)$ as the set of inverse images of
$\omega$ under $\mathcal{C}^{\ast}$ in the space $C(B_{K})^{\ast}$ of linear
functionals,
\begin{align}
(\mathcal{C}^{\ast})^{-1}(\omega):=  &  \{\nu\in C(B_{K})^{\ast};\nu=\nu
_{+}-\nu_{-},\nu_{\pm}\geq0,\nonumber\\
&  \nu_{-}\upharpoonright_{\mathcal{C}(\mathcal{S}_{x})}=0,\mathcal{C}^{\ast
}(\nu_{+})\upharpoonright_{\mathcal{S}_{x}}=\omega\upharpoonright
_{\mathcal{S}_{x}}\},
\end{align}
we can put Eq.(\ref{S-thermal}) into the similar form to Eq.(\ref{adjunction1}%
) as

\begin{itemize}
\item[iv)] The characterization and local thermal interpretation of a
non-equilibrium local state:

\begin{proposition}
\cite{Oji03} The following isomorphism holds for $\omega\in E_{x}$, $\rho
_{x}\in Th$ and a subspace $\mathcal{S}_{x}\subset\mathcal{T}_{x}$,%
\begin{equation}
E_{x}/\mathcal{S}_{x}(\omega,\mathcal{C}^{\ast}(\rho_{x}))\overset
{q\rightleftarrows c}{\simeq}Th/\mathcal{C}(\mathcal{S}_{x})((\mathcal{C}%
^{\ast})^{-1}(\omega),[\rho_{x}]), \label{adjunction2}%
\end{equation}
where $[\rho_{x}]:=\{\sigma\in Th;\sigma\upharpoonright_{\mathcal{C}%
(\mathcal{S}_{x})}=\rho_{x}\upharpoonright_{\mathcal{C}(\mathcal{S}_{x})}\}$.
The existence of $\rho_{x}$ to make the sets of arrows non-empty is equivalent
to the $\mathcal{S}_{x}$-thermality of $\omega$.
\end{proposition}
\end{itemize}

This relation can be viewed as a form of \textquotedblleft\textit{hierarchized
zeroth law of local thermodynamics}\textquotedblright; the reason for
mentioning the \textquotedblleft zeroth law\textquotedblright\ here is due to
the implicit relevance of measuring processes of local thermal observables
validating the above equalities, which require the \textit{contacts of two
bodies}, measured object(s) and measuring device(s), in a local thermal
equilibrium, conditional on the chosen $\mathcal{S}_{x}$. The transitivity of
this contact relation just corresponds to the localized and hierarchized
version of the standard zeroth law of thermodynamics.

We can use the relation
\begin{align}
\exists\nu &  =\nu_{+}-\nu_{-}\in(\mathcal{C}^{\ast})^{-1}(\omega)\text{ with
}\nu_{-}=0\Longleftrightarrow(\mathcal{C}^{\ast})^{-1}(\omega)=\{\nu\}\subset
Th\nonumber\\
&  \Longleftrightarrow\omega\in K\Longleftrightarrow\text{[maximal choice of
}\mathcal{S}_{x}^{\prime}\text{ s.t. }\mathcal{C}^{\ast}(\nu_{+}%
)\upharpoonright_{\mathcal{S}_{x}^{\prime}}=\omega\upharpoonright
_{\mathcal{S}_{x}^{\prime}}\text{]}=\mathcal{T}_{x},
\end{align}
for specifying the extent to which a non-equilibrium $\mathcal{S}_{x}$-thermal
$\omega$ deviates from equilibria belonging to $K$ by the \textit{failure of
positivity} ($\nu_{-}\neq0$) and can also measure it by the \textit{maximal
size} of $\mathcal{S}_{x}^{\prime}$ within the hierarchy of subspaces
$\mathcal{S}_{x}^{\prime}$ in $\mathcal{T}_{x}$ such that $\mathcal{S}%
_{x}^{\prime}\supset\mathcal{S}_{x}$, $\nu_{-}\upharpoonright_{\mathcal{C}%
(\mathcal{S}_{x}^{\prime})}=0$ with all the possible choices of $\nu
\in(\mathcal{C}^{\ast})^{-1}(\omega)$: owing to the presence of $\nu_{-}$,
$\omega$ ceases to be $\mathcal{S}_{x}^{\prime}$-thermal when $\mathcal{S}%
_{x}^{\prime}$ is so enlarged that $\nu_{-}\upharpoonright_{\mathcal{C}%
(\mathcal{S}_{x}^{\prime})}=0$ is invalidated, which shows that $\omega$
shares with reference states in $K$ only gross thermal properties described by
smaller $\mathcal{S}_{x}^{\prime}$. In this sense, the hierarchy of
$\mathcal{S}_{x}^{\prime}$ in $\mathcal{T}_{x}$ should have a close
relationship with the thermodynamic hierarchy at various scales appearing in
the transitions between non-equilibrium and equilibrium controlled by certain
family of \textit{coarse graining} procedures. Thus, we see that our selection
criterion can give a characterization of states identifiable as
non-equilibrium ones and, at the same time, provide associated relevant
physical interpretations of the selected states in a systematic way.

The two goals of identifying non-equilibrium local states admitting local
thermal interpretation and of describing their specific thermodynamic
properties are solved simultaneously by the above selection criterion based
upon a \textit{localized and hierarchized form of the zeroth law} of
thermodynamics. In this framework, we can identify at least three different
kinds of sources of derivations of an $\mathcal{S}_{x}$-thermal
non-equilibrium local state $\omega\in E_{x}$ from the genuine equilibrium
states $\omega_{\beta}$ as

\begin{itemize}
\item[a)] \textit{spacetime dependence} of thermal parameters such as
temperature distributions $x\longmapsto\beta(x)$,

\item[b)] \textit{statistical fluctuations} of thermal parameters at $x$
described by probability distributions $d\rho_{x}(\beta)\in Th$,
\end{itemize}

\noindent and

\begin{itemize}
\item[c)] essential deviations of local states $\omega\in E_{x}$ from states
in $K$ expressed by the \textit{positivity-violating} term $\nu_{-}\neq0$ in
$\nu=\nu_{+}-\nu_{-}\in(\mathcal{C}^{\ast})^{-1}(\omega)\subset C(B_{K}%
)^{\ast}$ with $\nu_{-}\upharpoonright_{\mathcal{C}(\mathcal{S}_{x}%
)}=0,\mathcal{C}^{\ast}(\nu_{+})\upharpoonright_{\mathcal{S}_{x}}%
=\omega\upharpoonright_{\mathcal{S}_{x}}$.
\end{itemize}

\section{Reformulation of DHR-DR sector theory}

\subsection{Basic results of DHR-DR theory}

According to the discussion in the previous section, we now try to reformulate
the essence of the DHR-DR sector theory into a physically more understandable
form. As the mathematical essence of the theory itself is very sophisticated
and complicated, it is not our aim here to reproduce it faithfully, for which
purpose interested readers are advised to look into their original papers
starting from \cite{DR90, DR89}. Before taking our approach to it in Sec.3.2,
however, we need to introduce the basic ingredients and to summarize the most
essential results of the DHR-DR sector theory. The starting point of the
theory with \textit{localizable charges} \cite{DHR, Haag} is as follows:

\begin{itemize}
\item A net $\mathcal{K}\ni\mathcal{O}\longmapsto\mathfrak{A}(\mathcal{O})$ of
von Neumann algebras $\mathfrak{A}(\mathcal{O})$ of \textit{local observables}
is defined on the set, $\mathcal{K}:=\{(a+V_{+})\cap(b-V_{+});a,b\in
\mathbb{R}^{4}\}$, of all double cones in the Minkowski spacetime
$\mathbb{R}^{4}$; it is assumed to satisfy

\begin{itemize}
\item isotony: $\mathcal{O}_{1}\subset\mathcal{O}_{2}\Longrightarrow
\mathfrak{A}(\mathcal{O}_{1})\subset\mathfrak{A}(\mathcal{O}_{2})$, allowing
the global (or, quasi-local) algebra of observables $\mathfrak{A}:=C^{\ast}$-
$\underset{\mathcal{K\ni O\nearrow}\mathbb{R}^{4}}{\underset{\longrightarrow
}{\lim}}\mathfrak{A}(\mathcal{O})$ to be defined as the C*-inductive limit,

\item relativistic covariance under the action of the Poincar\'{e} group
$\mathcal{P}_{+}^{\uparrow}:=\mathbb{R}^{4}\rtimes L_{+}^{\uparrow}%
\ni(a,\Lambda)\longmapsto\alpha_{(a,\Lambda)}\in Aut(\mathfrak{A)}$,
$\alpha_{(a,\Lambda)}(\mathfrak{A}(\mathcal{O}))=\mathfrak{A}(\Lambda
(\mathcal{O)}+a)$, and

\item local commutativity (or locality for short): $[\mathfrak{A}%
(\mathcal{O}_{1}),\mathfrak{A}(\mathcal{O}_{2})]=0$ for $\mathcal{O}%
_{1},\mathcal{O}_{2}\in\mathcal{K}$ spacelike separated (i.e., $\forall
x\in\mathcal{O}_{1},\forall y\in\mathcal{O}_{2}$, $(x-y)^{2}<0$).
\end{itemize}

\item \textbf{DHR\ criterion}: A physically relevant state $\omega\in
E_{\mathfrak{A}}$(: the set of all states of $\mathfrak{A}$ defined as
normalized positive linear functionals on $\mathfrak{A}$) around a pure vacuum
$\omega_{0}\in E_{\mathfrak{A}}$ is selected by the Doplicher-Haag-Roberts
(DHR)\textit{\ }criterion\footnote{In view of the local normality, this
criterion can be imposed on \textit{any }$\mathcal{O}\in\mathcal{K}$ as seen
in \cite{DR90}, but, we adopt this original form presented in \cite{DHR} in
relation to the notion of support of $\rho$ in the next item.} which requires
the GNS reprepresentation $\pi_{\omega}$ corresponding to $\omega$ to be
unitarily equivalent to the vacuum representation $\pi_{\omega_{0}}=:\pi_{0}$
in spacelike distance; i.e., $\exists\mathcal{O\in K}$ s.t. for $\forall
a\in\mathbb{R}^{4}$ with $\mathcal{O}_{a}:=\mathcal{O}+a\in\mathcal{K}$
\begin{equation}
\pi_{\omega}\upharpoonright_{\mathfrak{A}(\mathcal{O}_{a}^{\prime})}\cong
\pi_{0}\upharpoonright_{\mathfrak{A}(\mathcal{O}_{a}^{\prime})},
\label{DHRcrit}%
\end{equation}
where $\mathcal{O}^{\prime}:=\{x\in\mathbb{R}^{4}$; $(x-y)^{2}<0$ for $\forall
y\in\mathcal{O}\}$ is the causal complement of $\mathcal{O}$ and
$\mathfrak{A}(\mathcal{O}^{\prime}):=C^{\ast}$-$\underset{\mathcal{K\ni O}%
_{1}\subset\mathcal{O}^{\prime}}{\underset{\longrightarrow}{\lim}}%
\mathfrak{A}(\mathcal{O}_{1})$.

\item \textbf{Local endomorphisms}: In the GNS representation $(\pi
_{0},\mathfrak{H}_{0})$ corresponding to $\omega_{0}$, the validity of Haag
duality,
\begin{equation}
\pi_{0}(\mathfrak{A}(\mathcal{O}^{\prime}))^{\prime}=\pi_{0}(\mathfrak{A}%
(\mathcal{O}))^{\prime\prime},
\end{equation}
is assumed. On the basis of the standard postulates \cite{DHR}, the selection
criterion (\ref{DHRcrit}) can be shown to be equivalent to the existence of a
\textit{local endomorphism} $\rho\in End(\mathfrak{A})$ such that $\pi
_{\omega}=\pi_{0}\circ\rho$, localized in some $\mathcal{O}\in\mathcal{K}$ in
the sense of%
\begin{equation}
\rho(A)=A\text{ \ \ \ for }\forall A\in\mathfrak{A(}\mathcal{O}^{\prime}).
\end{equation}
In this situation, we say (in a rather sloppy way) that the \textit{support of
}$\rho$ is (contained in) $\mathcal{O}$: supp$(\rho)\subset\mathcal{O}$. Note
that an endomorphism $\rho$ preserves all the algebraic structure on
$\mathfrak{A}$ but that its image set $\rho(\mathfrak{A})$ can be strictly
smaller than $\mathfrak{A}$, $\rho(\mathfrak{A})\subsetneqq\mathfrak{A}$ which
is possible only for an infinite-dimensional algebra $\mathfrak{A}$.

\item \textbf{Transportability} (of charges associated with an
\textit{internal }symmetry): The above spacetime dependence of $\rho$ coming
from its localization region $\mathcal{O}$ can be absorbed into its
\textit{transportability}, namely, for any translation $a\in\mathbb{R}^{4}$,
there exists $\rho_{a}\in End(\mathfrak{A})$ with support in $\mathcal{O}+a$
and $\rho\cong\rho_{a}=Ad(u_{a})\circ\rho$ with a unitary $u_{a}%
\in\mathfrak{A}$. We denote
\begin{equation}
\Delta(\mathcal{O}):=\{\rho\in End(\mathfrak{A});\rho\text{: transportable and
localized in }\mathcal{O}\}.
\end{equation}

\item \textbf{DR-category }\cite{DR89}: Then a \textbf{C*-tensor category}
$\mathcal{T}$ which we call here a DR-category is defined as a full
subcategory of $End(\mathfrak{A})$ consisting of \textit{objects} $\rho
\in\Delta:=\cup_{\mathcal{O}\in\mathcal{K}}\Delta(\mathcal{O})$ and with
\textit{morphisms} (or, \textit{arrows}) given by intertwiners $T\in
\mathfrak{A}$ between $\rho,\sigma\in\Delta$ s.t. $T\rho(A)=\sigma(A)T$.
$\mathcal{T}$ has the \textit{permutation symmetry} due to the locality, and
is closed under \textit{direct sums} and \textit{subobjects} (due to the
Property B following from the spectrum condition, locality and weak
additivity) \cite{DR90}.
\end{itemize}

As promised in Sec.2, we encounter here a category $\mathcal{T}$, a
mathematical notion more general than a groupoind (corresponding to an
equivalence relation) in that its arrows are not necessarily invertible. By
$\mathcal{T}$ being a \textit{C*-tensor category} we mean

i) [\textit{C*-category}]: all the sets $\mathcal{T}(\rho,\sigma)$ of arrows
in $\mathcal{T}$ are Banach spaces over complex numbers $\mathbb{C}$ such that
a hermitian conjugation $\mathcal{T}(\rho,\sigma)\ni T\longmapsto T^{\ast}%
\in\mathcal{T}(\sigma,\rho)$ is so defined that the C*-norm property
$\left\vert \left\vert T^{\ast}T\right\vert \right\vert =\left\vert \left\vert
T\right\vert \right\vert ^{2}$ holds for the norms of arrows (which is
straightforward from $T\in\mathcal{T}(\rho,\sigma)\subset\mathfrak{A}$:
C*-algebra), and

ii) [\textit{tensor category}]: a tensor-product structure is defined on the
set of objects $\rho,\sigma,\cdots$, etc., by $\rho\otimes\sigma:=\rho\sigma$
and also on that of arrows. $S\otimes T$ is defined for $S\in\mathcal{T}%
(\rho_{1},\rho_{2}),T$ $\in\mathcal{T}(\sigma_{1},\sigma_{2})$ by $S\otimes
T:=S\rho_{1}(T)=\rho_{2}(T)S$, and satisfies $S\rho_{1}(T)\rho_{1}\sigma
_{1}(A)=\rho_{2}(\sigma_{2}(A)T)S=\rho_{2}\sigma_{2}(A)S\rho_{1}(T)$, which
means $S\otimes T\in\mathcal{T}(\rho_{1}\otimes\sigma_{1},\rho_{2}%
\otimes\sigma_{2})=\mathcal{T}(\rho_{1}\sigma_{1},\rho_{2}\sigma_{2})$ and
also $(S_{1}\otimes T_{1})(S_{2}\otimes T_{2})=S_{1}S_{2}\otimes T_{1}T_{2}$.

According to i), a C*-category with only one object is just a C*-algebra as a
Banach space equipped with product structure and the C*-norm. What is
remarkable is the tensor structure ii) which is shared by the category
$Rep_{G}$ of unitary representations $(\gamma,V_{\gamma})$ of a group $G$
(i.e., $\gamma(g)$: unitary operators in the inner product space $V_{\gamma}$
s.t. $\gamma(g_{1}g_{2})=\gamma(g_{1})\gamma(g_{2}),\gamma(g^{-1}%
)=\gamma(g)^{\ast},\gamma(e)=Id_{V_{\gamma}}$) whose arrows are intertwiners
between pairs of such representations, i.e., $T\in Rep_{G}(\gamma_{1}%
,\gamma_{2})\Longleftrightarrow T\gamma_{1}(g)=\gamma_{2}(g)T$ for $\forall
g\in G$.

In a word, the mathematical essence of Doplicher-Roberts theory is to verify
that, in spite of its abstract form as a certain category of local
endomorphisms $\rho$ on the observable algebra $\mathfrak{A}$, the
DR-category\textbf{ }$\mathcal{T}$ determined by the DHR criterion for
relevant states is isomorphic to the category $Rep_{G}$ of group
representations with a certain uniquely determined group $G$ to be identified
with the gauge group (of the 1st kind). Up to the technical details, the
essential contents can be summarized in the following basic results due to the
structure of $\mathcal{T}$ as a C*-tensor category\ having the
\textit{permutation symmetry}, \textit{direct sums}, \textit{subobjects} and
\textit{conjugates}:

\begin{itemize}
\item Unique existence of an \textit{internal symmetry group} $G$ such that
\begin{equation}
\mathcal{T}\simeq Rep_{G}\underset{\text{ \ \ \ \ \ \ \ \ \ \ \ \ \ \ \ \ }%
}{\overset{\text{Tannaka-Krein duality}}{\longleftrightarrow}}G=End_{\otimes
}(V),
\end{equation}
where $End_{\otimes}(V)$ is defined as the group of natural unitary
transformations $g=(g_{\rho})_{\rho\in\mathcal{T}}:V\overset{\cdot
}{\rightarrow}V$ from the C*-tensor functor $V:\mathcal{T}\hookrightarrow
Hilb$ to itself \cite{DR89, MacL} as characterized by $g_{\rho\sigma}=g_{\rho
}\otimes g_{\sigma}$ and the commutativity $Tg_{\rho_{1}}=g_{\rho_{2}}T$ of
the diagram:
\begin{equation}%
\begin{array}
[c]{ccccc}%
\rho_{1} &  & V_{\rho_{1}} & \overset{g_{\rho_{1}}}{\rightarrow} & V_{\rho
_{1}}\\
T\downarrow &  & T\downarrow & \circlearrowleft & \downarrow T\\
\rho_{2} &  & V_{\rho_{2}} & \overset{g_{\rho_{2}}}{\rightarrow} & V_{\rho
_{2}}%
\end{array}
.
\end{equation}
Here, $V$ embeds $\mathcal{T}$ into the category $Hilb$ of Hilbert spaces and
its image turns out to be just the category $Rep_{G}$ of unitary
representations $(\gamma,V_{\gamma})$ of a compact Lie group $G\subset SU(d)$
(owing to the presence of \textit{conjugates} in $\mathcal{T}$), where the
dimensionality$\ d$ is intrinsically defined in $\mathcal{T}$ by the
generating element $\rho\in\mathcal{T}$ \cite{DR89}\footnote{For the unique
existence of $G$, the functor $V$ should be so chosen that it maps the
`bosonized' form \cite{DR90} of permutation symmetry of $\mathcal{T}$ onto the
unique permutation symmetry of $Hilb$, as emphasized by Prof. Roberts.}. In
this formulation, the essence of Tannaka-Krein duality \cite{TK} is found in
the one-to-one correspondence,
\begin{equation}
_{\mathcal{I}}\backslash^{\mathcal{\Delta}}\ni\lbrack\rho]=[\rho_{\gamma
}]\longleftrightarrow\gamma=\gamma_{\rho}\in\hat{G}, \label{G-charge}%
\end{equation}
with $\rho\in\mathcal{\Delta}$ satisfying $\rho(\mathfrak{A})^{\prime}%
\cap\mathfrak{A}=\mathbb{C}\mathbf{1}$ (corresponding to the
\textit{irreducibility} of $\gamma_{\rho}$), and the identification $g_{\rho
}=\gamma_{\rho}(g)$ for $g\in G$, where $_{\mathcal{I}}\backslash
^{\mathcal{\Delta}}$ \ is the set of equivalence classes $\{Ad(v)\circ
\rho;\mathcal{O}\in\mathcal{K},v\in\mathfrak{A}(\mathcal{O})\}\subset\Delta$
of $\rho$ w.r.t. the action of inner automorphism group $\mathcal{I}%
=\{Ad(v);\mathcal{O}\in\mathcal{K},v\in\mathcal{U}(\mathfrak{A}(\mathcal{O}%
))$: unitary operators$\}$ and the group dual $\hat{G}$ is defined by the
totality of equivalence classes of continuous unitary irreducible
representations of $G$. \newline Once these are known, the relation
$g_{\rho\sigma}=g_{\rho}\otimes g_{\sigma}$ can be understood as
representating the tensor structure of representations $\gamma_{\rho}$ of $G$
(i.e., a representation of representations, which is sometime expressed by the
word \textquotedblleft birepresentation\textquotedblright\ corresponding to
the bidual $\hat{\hat{\Gamma}}$ of an abelian group $\Gamma$ isomorphic to
$\Gamma$ itself by the Pontryagin duality) and the relation $Tg_{\rho_{1}%
}=g_{\rho_{2}}T$ rewritten by $T\gamma_{\rho_{1}}(g)=\gamma_{\rho_{2}}(g)T$
simply shows that the intertwiner $T$ from an endomorphism $\rho_{1}$ to
another such $\rho_{2}$ is just the one from $\gamma_{\rho_{1}}$ to
$\gamma_{\rho_{2}}$ in the context of group representations.

\begin{remark}
The appearance of group structure here is due to the permutation symmetry
encoded in $\mathcal{T}$ coming from the local commutativity in the four
dimensional spacetime. In the two dimensional case, the permutation symmetry
is to be replaced by the braid group symmetry, as a consequence of which
quantum group symmetry arises instead of the familiar group (see
\cite{FroKer93} for wide perspectives of the relevant problems involving
\textquotedblleft quantum categories\textquotedblright).
\end{remark}

\item Unique existence of a \textit{field algebra }such that
\begin{align}
\mathfrak{F}:=\mathfrak{A}\underset{\mathcal{O}_{d}^{G}}{\otimes}%
\mathcal{O}_{d}  &  \text{ \ \ \ }\curvearrowleft G=Aut_{\mathfrak{A}%
}(\mathfrak{F})=Gal(\mathfrak{F}/\mathfrak{A})\\
&  :=\{\tau\in Aut(\mathfrak{F});\tau(A)=A,\text{ }\forall A\in\mathfrak{A}%
\}\text{ (: Galois group),}\nonumber
\end{align}
with $\mathfrak{A}=\mathfrak{F}^{G}$ (fixed-point algebra), where
$\mathcal{O}_{d}$ is the Cuntz algebra \cite{Cuntz} defined as the unique
simple C*-algebra generated by a $d$-dimensional Hilbert space $h_{d}$ of $d$
isometries $\psi_{i}$, $i=1,2,\cdots,d$,
\begin{equation}
\psi_{i}^{\ast}\psi_{j}=\delta_{ij}\mathbf{1,}\text{ \ \ \ }\sum_{i=1}^{d}%
\psi_{i}\psi_{i}^{\ast}=\mathbf{1,}%
\end{equation}
whose fixed-point subalgebra $\mathcal{O}_{d}^{G}$ is embedded into
$\mathfrak{A}$, $\mu:\mathcal{O}_{d}^{G}\hookrightarrow\mathfrak{A}$,
satisfying the relation $\mu\circ\sigma=\rho\circ\mu$ with respect to the
canonical endomorphism $\sigma$ of $\mathcal{O}_{d}$: $\sigma(C):=\sum
_{i=1}^{d}\psi_{i}C\psi_{i}^{\ast}$ for $C\in$ $\mathcal{O}_{d}$. As a linear
space, $\mathfrak{F}$ is uniquely defined as a tensor product of
$\mathfrak{A}$ as a right $\mathcal{O}_{d}^{G}$-module via $\mu$ and of
$\mathcal{O}_{d}$ as a left $\mathcal{O}_{d}^{G}$-module, and its product
structure is defined \cite{DR89} by%
\begin{align}
&  (A_{1}\underset{\mathcal{O}_{d}^{G}}{\otimes}\psi_{i_{1}}\cdots\psi_{i_{r}%
}\psi_{j_{1}}^{\ast}\cdots\psi_{j_{s}}^{\ast})(A_{2}\underset{\mathcal{O}%
_{d}^{G}}{\otimes}C)\nonumber\\
&  =[(-1)^{d-1}\sqrt{d}]^{s}A_{1}\rho^{r}(\underset{s}{\underbrace{R^{\ast
}\rho^{d-1}(\cdots(R^{\ast}\rho^{d-1}}}(A_{2}))\cdots)\nonumber\\
&  \text{ \ \ \ }\underset{\mathcal{O}_{d}^{G}}{\otimes}\psi_{i_{1}}\cdots
\psi_{i_{r}}\hat{\psi}_{j_{1}}\cdots\hat{\psi}_{j_{s}}C
\end{align}
for $A_{i}\in\mathfrak{A}$, $\psi_{i}\in h_{d}$, $C\in\mathcal{O}_{d}$, where
\begin{equation}
\hat{\psi}_{i}=1/\sqrt{(d-1)!}\sum_{p\in\mathbb{P}_{d}(i)}sgn(p)\psi
_{p(2)}\cdots\psi_{p(d)}%
\end{equation}
with $\mathbb{P}_{d}(i)$ the subset of permutations $p$ of $1,2,\cdots,d$ s.t.
$p(1)=i$ and $R=1/\sqrt{d}\mu(\sum_{i=1}^{d}\psi_{i}\hat{\psi}_{i}%
)\in\mathcal{T}(\iota,\rho^{d}).$ (For the unique existence of C*-norm see
\cite{DR89}.)

\item The local net structure of $\mathfrak{F}$ is provided consistently by
the local W*-algebras $\mathfrak{F}(\mathcal{O})$ generated from the family of
Hilbert spaces $H_{\rho}$, $\rho\in\Delta(\mathcal{O})$, in $\mathfrak{F}$,%
\begin{equation}
H_{\rho}:=\{\psi\in\mathfrak{F};\psi A=\rho(A)\psi\text{ for }\forall
A\in\mathfrak{A\}\subset F},
\end{equation}

whose inner product structure is due to the basic structural relation
$\mathfrak{A}^{\prime}\cap\mathfrak{F}=\mathbb{C}\mathbf{1}$ \cite{DR88, DR90}
(equivalent to the condition for all $G$-representations to be contained in
$\mathfrak{F}$). Mathematically, the uniqueness of $\mathfrak{F}$ and of
$\mathfrak{F}(\mathcal{O})$ comes from the fact that they are the solutions of
the \textit{universality} problem to make the following diagram commutative,
which automatically ensures the uniqueness and consistency of the
constructions of $\mathfrak{F}$ and $\mathfrak{F}(\mathcal{O})$ from
$\mathfrak{A}$ and $\mathfrak{A}(\mathcal{O})$, respectively:
\begin{equation}%
\begin{array}
[c]{ccc}%
\mathfrak{A} & \hookrightarrow & \mathfrak{F}\\
\rotatebox{90}{$\hookrightarrow$} &  & \rotatebox{90}{$\hookrightarrow$}\\
\mathfrak{A}(\mathcal{O}) & \hookrightarrow & \mathfrak{F}(\mathcal{O})\\
{\mu_{\mathcal{O}}}\rotatebox{90}{$\hookrightarrow$}\text{\ \ \ \ } &  &
\text{ \ \ }\rotatebox{90}{$\hookrightarrow$}\zeta_{\mathcal{O}}\\
\mathcal{O}_{d}^{G} & \hookrightarrow & \mathcal{O}_{d}%
\end{array}
.
\end{equation}

\item The \textit{sector structure} in the irreducible vacuum representation
$(\pi,\mathfrak{H})$ of the constructed field algebra $\mathfrak{F}$ is
understood as follows: first, the group $G$ of symmetry arising in this way is
\textbf{unbroken} with a unitary implementer $U:G\rightarrow\mathcal{U}%
(\mathfrak{H}),$ $\pi(\tau_{g}(F))=U(g)\pi(F)U(g)^{\ast}$ and is
\textit{global }(i.e., gauge symmetry of the 1st kind) [due to the
transportability in spacetime imposed on each $\rho\in\mathcal{T}$]. This
representation is realized as the induced representation of $\mathfrak{F}$
from the pure vacuum representation $(\pi_{0},\mathfrak{H}_{0})$ of
$\mathfrak{A}$ through the conditional expectation of $G$-average
$m:\mathfrak{F\rightarrow A}$ defined by
\begin{equation}
\mathfrak{F}\ni F\longmapsto m(F):=\int_{G}dg\tau_{g}(F)\in\mathfrak{A,}%
\end{equation}
arising from the vacuum state $\bar{\omega}$ of $\mathfrak{F}$ given by
$\bar{\omega}(F):=\omega_{0}(m(F))$, $\pi=\pi_{\bar{\omega}}$, $\mathfrak{H=H}%
_{\bar{\omega}}$. Then, $\mathfrak{H}$ contains the starting Hilbert space
$\mathfrak{H}_{0}$ of the vacuum representation $\pi_{0}$ of $\mathfrak{A}$ as
a cyclic $G$-fixed-point subspace, $\mathfrak{H}_{0}=\mathfrak{H}^{G}=\{\xi
\in\mathfrak{H}$; $U(g)\xi=\xi$ for $\forall g\in G\}$, $\overline
{\pi(\mathfrak{F})\mathfrak{H}_{0}}=\mathfrak{H}$. Then $\mathfrak{H}$ is
decomposed into a direct sum in the following form \cite{DHR},
\begin{align}
\mathfrak{H}  &  =\underset{\gamma\in\hat{G}}{\oplus}(\mathfrak{H}_{\gamma
}\otimes V_{\gamma}),\\
\pi(\mathfrak{A})  &  =\underset{\gamma\in\hat{G}}{\oplus}(\pi_{\gamma
}(\mathfrak{A})\otimes\mathbf{1}_{V_{\gamma}}),\text{ \ \ \ }U(G)=\underset
{\gamma\in\hat{G}}{\oplus}(\mathbf{1}_{\mathfrak{H}_{\gamma}}\otimes
\gamma(G)), \label{sector}%
\end{align}
where \textbf{superselection sectors} defined as equivalence classes of
irreducible representations $(\pi_{\gamma},\mathfrak{H}_{\gamma})$ of
$\mathfrak{A}$ are \textit{in one-to-one correspondence}, $\pi_{\gamma}%
=\pi_{0}\circ\rho_{\gamma}\longleftrightarrow\lbrack\rho_{\gamma}%
]\in_{\mathcal{I}}\backslash^{\mathcal{\Delta}}$ $\longleftrightarrow
(\gamma,V_{\gamma})$, with equivalence classes of irreducible unitary
representations $(\gamma,V_{\gamma})\in\hat{G}$ of $G$.
\end{itemize}

\subsection{Centre and central decompositions}

What is important about (\ref{sector}) is the existence of a
\textit{non-trivial centre} of $\pi(\mathfrak{A})^{\prime\prime}$,
\begin{align}
\mathfrak{Z}_{\pi}(\mathfrak{A})  &  :=\mathfrak{Z}(\pi(\mathfrak{A}%
)^{\prime\prime})=\pi(\mathfrak{A})^{\prime\prime}\cap\pi(\mathfrak{A}%
)^{\prime}=\mathfrak{Z}(U(G)^{\prime\prime})\nonumber\\
&  =\underset{\gamma\in\hat{G}}{\oplus}\mathbb{C}(\mathbf{1}_{\mathfrak{H}%
_{\gamma}}\otimes\mathbf{1}_{V_{\gamma}})=l^{\infty}(\hat{G}), \label{centre}%
\end{align}
which implies that points $\gamma\in\hat{G}$ or (generalized) observables
$(f_{\gamma})_{\gamma\in\hat{G}}\in l^{\infty}(\hat{G})$ belonging to the
centre of $\pi(\mathfrak{A})^{\prime\prime}$ as $G$-invariants are
\textit{order parameters }to distinguish among different sectors carrying
different $G$-representations (in parallel with the similar role of Casimir
operators in the enveloping algebra of Lie algebra $\mathfrak{g}$). From our
viewpoint, the physical essence of the long and complicated mathematical story
involved in the DHR-DR sector theory can be summarized as follows: a pure
state $\omega\in E_{\mathfrak{A}}$ of the observable algebra $\mathfrak{A}$ is
characterized as one carrying a localized charge by the DHR selection
criterion, Eq.(\ref{DHRcrit}), for $\pi_{\omega}\in Rep_{\mathfrak{A}}$, which
is equivalent to the existence of $\rho\in\mathcal{T}$[: DR category ($\subset
End(\mathfrak{A})$)] s.t. $\pi_{\omega}=\pi_{0}\circ\rho$. Via the
Doplicher-Roberts categorical equivalence $\mathcal{T}\simeq Rep_{G}$, this
data is further transformed into a $G$-charge $\gamma=\gamma_{\rho}\in\hat
{G}\subset Rep_{G}$ describing the $G$-behaviour of the state $\omega\circ m$
of the field algebra $\mathfrak{F}$ induced from $\mathfrak{A}$ through the
conditional expectation $m$, as a result of which the sector structure of
states of $\mathfrak{A}$ selected by the DHR-criterion (DHR-selected states
for short) is parametrized and classified by $Spec(\mathfrak{Z}_{\pi
}(\mathfrak{A}))\simeq\hat{G}$. Namely, we can draw such a flow chart:

a DHR-selected state $\omega\in E_{\mathfrak{A}}$ $\overset{\text{GNS-rep.}%
}{\Longrightarrow}$ [$\pi_{\omega}\in\{\pi_{0}\circ\rho;\rho\in\mathcal{T}%
\}(\subset Rep\mathfrak{A})$]\newline$\overset{\text{DHR}}{\Longleftrightarrow
}$ [$\rho\in\mathcal{T}(\subset End(\mathfrak{A}))\overset{\text{DR}}{\simeq
}Rep_{G}$] $\Longleftrightarrow$ [$\gamma_{\rho}\in\hat{G}(\subset Rep_{G}%
)$]\newline$\Longrightarrow$ [sectors of $\mathfrak{A}$ parametrized by
$Spec(\mathfrak{Z}_{\pi}(\mathfrak{A}))\simeq\hat{G}$ in the irreducible
vacuum representation $(\pi,\mathfrak{H})$ of $\mathfrak{F}$].

While the similarity to the scheme in Sec.2 starts now to emerge, we note that
the relation of the mathematical notion of \textit{representations} to the
actual physical situations is rather indirect in comparison to that of
\textit{states}, in view of which it is desirable to reformulate the above
scheme into such a form that the parallelism with Sec.2 becomes more evident.
So, we need to examine here as to how one can physically attain the
information on the $G$-charge contents of a given state $\omega$ of
$\mathfrak{A}$ encoded in $\mathfrak{Z}_{\pi}(\mathfrak{A})$ as in
Eq.(\ref{centre}), which has not been discussed in the traditional context of
the sector theory. For this purpose, starting from a generic mixture $\omega$
of DHR-selected states, we aim at an expression for it of
\textit{Fourier-decomposition type} similar to Eq.(\ref{Decomp}),
$\omega_{\rho}=\int_{B_{K}}d\rho(\beta,\mu)\omega_{\beta,\mu}=\mathcal{C}%
^{\ast}(\rho)$, for a thermal reference state $\omega_{\rho}\in K$ in Sec.2.1.

We consider now the mutual relation between states and representations of
$\mathfrak{A}$. In the direction from states to representations, the GNS
construction, $E_{\mathfrak{A}}\ni\omega\overset{\text{GNS}}{\longmapsto}%
(\pi_{\omega},\mathfrak{H}_{\omega},\Omega_{\omega})$ s.t. $\omega
(A)=\langle\Omega_{\omega}\ |\ \pi_{\omega}(A)\Omega_{\omega}\rangle$,
$\mathfrak{H}_{\omega}=\overline{\pi_{\omega}(\mathfrak{A})\Omega_{\omega}}$
induces a canonical map $E_{\mathfrak{A}}\ni\omega\longmapsto(\pi_{\omega
},\mathfrak{H}_{\omega})\in Rep_{\mathfrak{A}}$ (well-defined up to unitary
equivalence). The opposite direction, however, involves the inevitable
many-valuedness which necessitates the treatment of suitable sets of states,
for instance, the set of vector states, $(\eta,\mathfrak{H}_{\eta}%
)\longmapsto\mathfrak{V}_{\eta}\equiv\{\omega_{\Psi}\in E_{\mathfrak{A}%
};\omega_{\Psi}(A):=\langle\Psi\ |\ \eta(A)\Psi\rangle$ for $\forall
A\in\mathfrak{A}$, $\Psi\in\mathfrak{H}_{\eta}\}$, or that of density-matrix
states in $(\eta,\mathfrak{H}_{\eta})$. Since the latter choice has a natural
connection with the von Neumann algebra $\eta(\mathfrak{A})^{\prime\prime}$ of
the representation $\eta$, it has a name, a \textit{folium}\footnote{A folium
$\mathfrak{f}(\eta)$ is related with the von Neumannn algebra $\eta
(\mathfrak{A})^{\prime\prime}$ in such a way that its linear hull
$Lin(\mathfrak{f}(\eta))$ consisting of linear combinations of states in
$\mathfrak{f}(\eta)$ is the \textit{predual} $\eta(\mathfrak{A})_{\ast
}^{\prime\prime}$\textit{ }of $\eta(\mathfrak{A})^{\prime\prime}$,
$Lin(\mathfrak{f}(\eta))=\eta(\mathfrak{A})_{\ast}^{\prime\prime}$ uniquely
characterized by the relation $(\eta(\mathfrak{A})_{\ast}^{\prime\prime
})^{\ast}=\eta(\mathfrak{A})^{\prime\prime}$.} associated to $(\eta
,\mathfrak{H}_{\eta})$, which we denote by
\begin{equation}
\mathfrak{f}(\eta):=\{\omega\in E_{\mathfrak{A}};\exists\sigma:\text{density
operator in }\mathfrak{H}_{\eta}\text{ s.t. }\omega(A)=Tr_{\mathfrak{H}%
}[\sigma\eta(A)]\},
\end{equation}
and is also related to a state $\omega$ by $\mathfrak{f}(\omega):=\mathfrak{f}%
(\pi_{\omega})$ by means of the corresponding GNS representation $\pi_{\omega
}$. A state in $\mathfrak{f}(\eta)$ is also called a $\eta$-normal state of
$\mathfrak{A}$.

Then a state $\omega\in E_{\mathfrak{A}}$ of $\mathfrak{A}$ is a mixture of
DHR-selected states if and only if $\omega\in\mathfrak{f}(\pi)$ (with $\pi$
the restriction to $\mathfrak{A}$ of the vacuum representation $(\pi
,\mathfrak{H})$ of $\mathfrak{F}$ induced from the vacuum representation
$(\pi_{0},\mathfrak{H}_{0})$ of $\mathfrak{A}$), which is also equivalent to
the existence of an extension $\tilde{\omega}$ of $\omega$ to the von Neumann
algebra $\pi(\mathfrak{A})^{\prime\prime}$ given by $\tilde{\omega}(\tilde
{A})=Tr_{\mathfrak{H}}(\sigma_{\omega}\tilde{A})$ for $\tilde{A}\in
\pi(\mathfrak{A})^{\prime\prime}$ with such a density operator $\sigma
_{\omega}$ in $\mathfrak{H}$ that $\omega(A)=Tr_{\mathfrak{H}}(\sigma_{\omega
}\pi(A))$. Through the \textit{central decomposition} for the
\textquotedblleft simultaneous diagonalization\textquotedblright\ of centre
$\mathfrak{Z}_{\pi}(\mathfrak{A})=l^{\infty}(\hat{G})$, such a state
$\omega\in\mathfrak{f}(\pi)$ can be uniquely decomposed into the sum of factor
states $\omega_{\gamma}$ corresponding to $\gamma\in\hat{G}$:
\begin{equation}
\omega(A)=\sum_{\gamma\in\hat{G}}\mu_{\omega}(\gamma)\omega_{\gamma}(A).
\label{cenral_decomp}%
\end{equation}
Thus, we have a \textit{q}$\rightarrow$\textit{c channel }$\omega
\longmapsto\mu_{\omega}$ transforming quantum states into probability
distributions over the spectrum $\hat{G}$ of $\mathfrak{Z}_{\pi}%
(\mathfrak{A})$, which describes $G$-charge contents of each such quantum
state $\omega\in\mathfrak{f}(\pi)$ in terms of a probability distribution
$\mu_{\omega}=\{\mu_{\omega}(\gamma)\}_{\gamma\in\hat{G}}$ over$\ \hat{G}$.
This is in parallel with the integral decomposition Eq.(\ref{Decomp}).
However, one important difference should be noted here: within a sector
$(\pi_{\gamma},\mathfrak{H}_{\gamma})$ of the same $G$-charge $\gamma$, there
exist \textit{many} different states $\omega_{\gamma}$ showing different
behaviours under $\mathfrak{A}$, e.g., with different localization or
different energy-momentum spectrum, as energy-momentum (tensor) is invariant
under $G$.\footnote{This situation was carelessly overlooked in the original
version of this paper, as pointed out by Prof. J. E. Roberts, to whom I am
very grateful.}. Thus, in contrast to the thermal situation with fixed choice
of $\omega_{\beta,\mu}$, each factor state $\omega_{\gamma}$ appearing on the
right-hand side of Eq.(\ref{cenral_decomp}) may vary depending upon $\omega
\in\mathfrak{f}(\pi)$. In the former case, different factor KMS states
$\omega_{\beta,\mu}$ are always disjoint \cite{BR2} corresponding to different
order parameters (because of the uniqueness of a KMS state within its folium),
whereas what is shared in common by all the pure states $\omega_{\gamma}$
within a sector is just the unitary equivalence class $[\pi_{\omega_{\gamma}%
}]$ of the corresponding GNS \textit{representation} $\pi_{\omega_{\gamma}}$
of $\mathfrak{A}$ in terms of which all the above equivalent expressions
starting from the DHR criterion (\ref{DHRcrit}) are given. Since this point is
related to the equivalence of endomorphisms $\rho\cong Ad(u)\circ\rho$ for
$\rho\in\Delta$ w.r.t. $Ad(u)\in\mathcal{I}=Inn(\mathfrak{A})$ \cite{DHR}, we
should resolve this ambiguity to extract internal symmetry aspects of a given
\textit{state}. In view of the fact that local subalgebras $\mathfrak{A}%
(\mathcal{O})$ are \textit{factor} von Neumann algebras \textit{without
centres }from which the non-trivial centre $\mathfrak{Z}_{\pi}(\mathfrak{A})$
arises only in the weak closure $\pi(\mathfrak{A})^{\prime\prime}$ of the
\textit{global} algebra $\mathfrak{A}$, it is also interesting to ask a
related question as to how we can attain \textit{locally }and
\textit{minimally}\footnote{If we are allowed to collect and to introduce
\textit{all} the information concerning $\mathfrak{A}$, then the
\textquotedblleft ambiguities\textquotedblright\ trivially disappear, because
of their origins coming from the choices of states within a given sector and
from that of a representative $\rho$ among equivalent ones $Ad(u)\circ\rho$,
$Ad(u)\in Inn(\mathfrak{A})$.}\textit{ }the above solution in physical
situations according to the spirit of local quantum physics \cite{Haag}.

This is consistently achieved in use of $\rho\in\mathcal{T}$ as follows, in
parallel with the previous section. Choose a \textit{representative}
$\rho_{\gamma}$ from each equivalence class $[\rho_{\gamma}]\in_{\mathcal{I}%
}\backslash^{\mathcal{\Delta}}$, which amounts to a choice of a cross section
$\hat{G}\ni\gamma\longmapsto\rho_{\gamma}\in\lbrack\rho_{\gamma}]\subset
\Delta$ of a bundle $\Delta\twoheadrightarrow_{\mathcal{I}}\backslash
^{\mathcal{\Delta}}\simeq\hat{G}=Spec(\mathfrak{Z}_{\pi}(\mathfrak{A}))$.
After identifying a compact Lie group $G$, such a choice can be achieved,
e.g., by choosing one $\rho_{\gamma_{0}}$\ corresponding to the
\textit{fundamental representation} $\gamma_{0}$ of $G$; $\rho_{\gamma}$ for
arbitrary $\gamma\in\hat{G}$ can be extracted from $\rho_{\gamma_{0}}^{n}%
$\ with suitable $n\in\mathbb{N}$ as a direct-sum component, by means of
Clebsch-Gordan coefficients. In view of the physical meaning of $\rho$'s, this
choice can be interpreted as a specification of procedures to create
$G$-charges from the vacuum.

\subsection{Physical interpretation by conditional expectation as
\textit{c}$\rightarrow$\textit{q} channel and its \textquotedblleft
inverse\textquotedblright}

Then choosing an everywhere non-vanishing probability distribution $\mu_{G}$
over $\hat{G}$, $\mu_{G}=(\mu_{\gamma})_{\gamma\in\hat{G}}\in(0,1)^{\hat{G}}$,
$\sum_{\gamma\in\hat{G}}\mu_{\gamma}=1$, we can define a \textit{central
measure }$\mu$ on $E_{\mathfrak{A}}$ with support $\{\omega_{\gamma}%
:=\omega_{0}\circ\rho_{\gamma};\gamma\in\hat{G}\}$ in the state space
$E_{\mathfrak{A}}$ whose barycentre $\omega_{\mu}$ is given by
\begin{equation}
\omega_{\mu}(A):=\sum_{\gamma\in\hat{G}}\mu_{\gamma}\omega_{0}\circ
\rho_{\gamma}(A).
\end{equation}
This allows us also to define, in a similar way to the thermal situation, a
\textit{conditional expectation} $\Lambda_{\mu}:\mathfrak{A\rightarrow Z}%
_{\pi}(\mathfrak{A})$ as a \textit{c}$\rightarrow$\textit{q} channel s.t.
$\Lambda_{\mu}(A):=[\hat{G}\ni\gamma\longmapsto\omega_{0}\circ\rho_{\gamma
}(A)]\in\mathfrak{Z}_{\pi}(\mathfrak{A})$, $\Lambda_{\mu}^{\ast}(\nu
)(A)=\sum_{\gamma\in\hat{G}}\nu_{\gamma}[\Lambda_{\mu}(A)](\gamma
)=\sum_{\gamma\in\hat{G}}\nu_{\gamma}\omega_{0}\circ\rho_{\gamma}(A)$. Here
the definition of $\Lambda_{\mu}$ depends on the choice of a cross section
$\hat{G}\ni\gamma\longmapsto\rho_{\gamma}\in\lbrack\rho_{\gamma}]\subset
\Delta$ but is independent of the particular assignment of a probability
weight $\mu_{\gamma}$ to each $\gamma\in\hat{G}$. In use of this freedom we
see now that, similarly to the discussion in Sec.2.1, the central measure
$\mu$ as a \textit{q}$\rightarrow$\textit{c channel} allows physical
interpretation w.r.t. $G$ of all states of such forms as $\Lambda_{\mu}^{\ast
}(\nu)=\sum_{\gamma\in\hat{G}}\nu_{\gamma}\omega_{0}\circ\rho_{\gamma}\in
E_{\mathfrak{A}}$ with $\nu=(\nu_{\gamma})_{\gamma\in\hat{G}}\in M_{1}(\hat
{G}):=\{(\nu_{\gamma}^{\prime})_{\gamma\in\hat{G}};$ $\nu_{\gamma}^{\prime
}\geq0,\sum_{\gamma\in\hat{G}}\nu_{\gamma}^{\prime}=1\}$. Defining a map $W$
by
\begin{equation}
W:End(\mathfrak{A})\ni\rho\longmapsto\omega_{0}\circ\rho\in E_{\mathfrak{A}},
\end{equation}
we see the relations
\begin{align}
&  [\Lambda_{\mu}(A)](\gamma)=\omega_{0}\circ\rho_{\gamma}(A)=[W(\rho_{\gamma
})](A);\nonumber\\
&  \Lambda_{\mu}^{\ast}(\nu)(A)=\nu(\Lambda_{\mu}(A))=\sum_{\gamma\in\hat{G}%
}\nu_{\gamma}[\Lambda_{\mu}(A)](\gamma)=(\sum_{\gamma\in\hat{G}}\nu_{\gamma
}\omega_{0}\circ\rho_{\gamma})(A)\nonumber\\
&  \Longrightarrow\Lambda_{\mu}^{\ast}(\nu)=\sum_{\gamma\in\hat{G}}\nu
_{\gamma}\omega_{0}\circ\rho_{\gamma}=\sum_{\gamma\in\hat{G}}\nu_{\gamma
}W(\rho_{\gamma}). \label{charge}%
\end{align}
Therefore, the map $\Lambda_{\mu}^{\ast}$ extends $W$ to \textquotedblleft
convex combinations\textquotedblright\ of $\rho_{\gamma}$'s, and acts as a
\textquotedblleft\textit{charging} map\textquotedblright\ to create from the
vacuum $\omega_{0}$ a state $\Lambda_{\mu}^{\ast}(\nu)=\sum_{\gamma\in\hat{G}%
}\nu_{\gamma}(\omega_{0}\circ\rho_{\gamma})$ whose charge contents are
described by the charge distribution $\nu=(\nu_{\gamma})_{\gamma\in\hat{G}}\in
M_{1}(\hat{G})$ over the group dual $\hat{G}$. The role of the chosen cross
section $\gamma\longmapsto\rho_{\gamma}$ and the state family $E_{\mu
}:=\Lambda_{\mu}^{\ast}(M_{1}(\hat{G}))=\{\sum_{\gamma\in\hat{G}}\nu_{\gamma
}\omega_{0}\circ\rho_{\gamma};\nu_{\gamma}\geq0,\sum_{\gamma\in\hat{G}}%
\nu_{\gamma}=1\}\subset E_{\mathfrak{A}}$ is just to make the \textit{c}%
$\rightarrow$\textit{q} channel $\Lambda_{\mu}^{\ast}$ \textit{invertible }on
$E_{\mu}$, $E_{\mu}\ni\omega=\Lambda_{\mu}^{\ast}(\nu)\longmapsto\nu\in
M_{1}(\hat{G})$, to give a physical interpretation of $\omega$ w.r.t. $G$ in
terms of $\nu$.

As far as the internal symmetry aspect is concerned, we see that this setup is
already sufficient for providing any given state $\omega\in\mathfrak{f}(\pi)$
with its physical interpretation owing to the above observation and the simple
relation between central observables and folia: any states, $\varpi_{\gamma
}\in\mathfrak{f}(\omega_{\gamma})$, in a folium of the \textit{factorial}
state $\omega_{\gamma}=\omega_{0}\circ\rho_{\gamma}$ yield the same
expectation value $\varpi_{\gamma}(f)=f_{\gamma}$\ to each central observable
$f=(f_{\gamma})_{\gamma\in\hat{G}}\in l^{\infty}(\hat{G})=\mathfrak{Z}_{\pi
}(\mathfrak{A})$ which is \textquotedblleft diagonalized\textquotedblright\ in
the central decomposition. Therefore, we arrive at a similar formula to
Eq.(\ref{adjunction1}) in Sec.2.1 as

\begin{proposition}
Selection and interpretation of $G$-charges:
\begin{align}
&  (\mathfrak{f}(\pi)/\mathfrak{Z}_{\pi}(\mathfrak{A}))(\omega,\Lambda_{\mu
}^{\ast}(\nu))\simeq M_{1}(\hat{G})(\mu_{\omega},\nu)\nonumber\\
&  \Longleftrightarrow\mathfrak{f}(\omega)=\mathfrak{f}(\Lambda_{\mu}^{\ast
}(\nu))\Longleftrightarrow\mu_{\omega}(\gamma)=\nu_{\gamma}\text{ (for
}\forall\gamma\in\hat{G}\text{)}. \label{adjunction}%
\end{align}

\end{proposition}

To obtain a formula of Fourier-decomposition type similar to Eq.(\ref{Decomp}%
), however, we need to exhibit the additional elements appearing in the many
to one correspondence between states and representations [$E_{\mathfrak{A}}%
\ni\omega\overset{\text{GNS}}{\Longleftrightarrow}(\pi_{\omega},\mathfrak{H}%
_{\omega},\Omega_{\omega})\underset{\dashleftarrow}{\overset{\text{many to
one}}{\rightarrow}}(\pi_{\omega},\mathfrak{H}_{\omega})\in Rep_{\mathfrak{A}}%
$], in order to relate an arbitrary state $\phi=\sum_{\gamma\in\hat{G}}%
\nu_{\gamma}\varpi_{\gamma}\in\mathfrak{f}(\pi),$ $\varpi_{\gamma}%
\in\mathfrak{f}(\omega\circ\rho_{\gamma})$ to the family $E_{\mu}$. Since each
pure state belonging to $\mathfrak{f}(\omega\circ\rho_{\gamma})$ is written as
$\omega\circ\sigma_{\gamma}$ with $\sigma_{\gamma}$ related to $\rho_{\gamma}$
through $\sigma_{\gamma}(A)=u_{\gamma}^{\ast}\rho_{\gamma}(A)u_{\gamma}$,
$u_{\gamma}\in\mathcal{U}(\mathfrak{A})$, we have, for $\forall\phi
\in\mathfrak{f}(\pi)$ and $\forall A\in\mathfrak{A}$,
\begin{align}
\phi(A)  &  =\sum_{\gamma\in\hat{G}}\nu_{\gamma}\sum_{i\in I_{\gamma}}%
p_{i}^{\gamma}\omega_{0}\circ Ad(u_{\gamma,i}^{\ast})\circ\rho_{\gamma}(A)\\
&  =\sum_{\gamma\in\hat{G}}\nu_{\gamma}\sum_{i\in I_{\gamma}}p_{i}^{\gamma
}\langle u_{\gamma,i}\Omega_{0}\ |\ \pi_{0}\circ\rho_{\gamma}(A)u_{\gamma
,i}\Omega_{0}\rangle,
\end{align}
with $p_{i}^{\gamma}\in\lbrack0,1]$, $\sum_{i\in I_{\gamma}}p_{i}^{\gamma}=1$,
$u_{\gamma,i}\in\mathcal{U}(\mathfrak{A})$ for $\forall\gamma\in\hat{G}$,
$\forall i\in I_{\gamma}$. Here $\nu_{\gamma}$ is the probability to find the
sector with $G$-charge $\gamma\in\hat{G}$ in the state $\phi$ and
$p_{i}^{\gamma}$ is the conditional probability to find the state $\langle
u_{\gamma,i}\Omega_{0}\ |\ \pi_{0}\circ\rho_{\gamma}(-)u_{\gamma,i}\Omega
_{0}\rangle$ associated to the vector $u_{\gamma,i}\Omega_{0}$, knowing that
the system is already in the sector with $\gamma$.

Since the \textquotedblleft gap\textquotedblright\ between $u_{\gamma,i}%
\Omega_{0}$ and $\Omega_{0}$ is due to $u_{\gamma,i}\in\mathcal{U}%
(\mathfrak{A})$, its \textquotedblleft observabality\textquotedblright\ should
enable one to find some physical processes to identify it, for instance,
involving \textit{energy-momentum }(\textit{as observables}) by some limits of
taking the \textit{lowest energy state} among $\{u_{\gamma}\Omega
_{0};u_{\gamma}\in\mathcal{U}(\mathfrak{A})\}$, etc. (Actually this is the
same problem as discussed above concerning the choice of a section of
$\Delta\twoheadrightarrow_{\mathcal{I}}\backslash^{\mathcal{\Delta}}\simeq
\hat{G}$ in a different disguise. If we combine the data of relevant
observables, such as energy-momentum, from the beginning, this can be totally
absorbed into the choice of a section.) Once this is done, any other states
$\phi=\sum_{\gamma\in\hat{G}}\nu_{\gamma}\sum_{i\in I_{\gamma}}p_{i}^{\gamma
}\omega_{0}\circ Ad(u_{\gamma,i}^{\ast})\circ\rho_{\gamma}\in\mathfrak{f}%
(\pi)$ can be related to the corresponding $\Lambda_{\mu}^{\ast}(\nu)\in
E_{\mu}$ through the measurement of relevant observables (e.g., energy
momentum) and/or the limiting procedures to pick up $\Omega_{0}$ as the lowest
energy state among $u_{\gamma}\Omega_{0}$ with $u_{\gamma}\in\mathcal{U}%
(\mathfrak{A})$.

If $\phi=\sum_{\gamma\in\hat{G}}\nu_{\gamma}\varpi_{\gamma}\in\mathfrak{f}%
(\pi)$ has such a decomposition into sectors that its component factorial
states $\varpi_{\gamma}\in\mathfrak{f}(\omega\circ\rho_{\gamma})$ are all
\textit{pure}, there is a different but equivalent formulation in use of a
reducible representation $\gamma_{\nu}:=\oplus_{\gamma\in\hat{G},\nu_{\gamma
}\neq0}\gamma\in RepG$, which may look more familiar for treating the same
situation. For this purpose, we use the invariance of the vacuum state under
$U(G)$ which implies the following relations in terms of the conditional
expectation $m:\mathfrak{F\rightarrow A}=\mathfrak{F}^{G}$, $m(F)=\int
_{G}dg\tau_{g}(F)$:
\begin{equation}
(\omega_{0}\circ\rho_{\gamma})(m(F)))=\langle\Omega_{0}\ |\ \sum_{i}\psi
_{i}^{\gamma}m(F)\psi_{i}^{\gamma\ast}\Omega_{0}\rangle,
\end{equation}
where the last expression is understood in the representation space
$\mathfrak{H}$ of $\mathfrak{F}$ and $\psi_{i}^{\gamma}\in\mathfrak{F}$ are
such that $\psi_{i}^{\gamma}\pi(A)=\pi\circ\rho_{\gamma}(A)\psi_{i}^{\gamma}$
for $\forall A\in\mathfrak{A}$ (coming from the Cuntz algebra $\mathcal{O}%
_{d}$). Then owing to the disjointness among different sectors, the state
$\Lambda_{\mu}^{\ast}(\nu)$ can be rewritten as an induced state $\Lambda
_{\mu}^{\ast}(\nu)\circ m$ of $\mathfrak{F}$ by
\begin{align}
\Lambda_{\mu}^{\ast}(\nu)(m(F))  &  =\sum_{\gamma\in\hat{G}}\nu_{\gamma}%
\omega_{0}\circ\rho_{\gamma}(m(F))=\sum_{\gamma\in\hat{G}}\sum_{i}\langle
\sqrt{\nu_{\gamma}}\psi_{i}^{\gamma\ast}\Omega_{0}\ |\ m(F)\sqrt{\nu_{\gamma}%
}\psi_{i}^{\gamma\ast}\Omega_{0}\rangle\nonumber\\
&  =\langle\Psi\ |\ m(F)\Psi\rangle=\langle\Psi\ |\ F\Psi\rangle,
\end{align}
with a vector
\begin{equation}
\Psi:=\sum_{\gamma\in\hat{G}}\sum_{i}\sqrt{\nu_{\gamma}}\psi_{i}^{\gamma\ast
}\Omega_{0}\in\mathfrak{H} \label{rep_vec}%
\end{equation}
belonging to the above mentioned reducible representation $\gamma_{\nu
}:=\oplus_{\gamma\in\hat{G},\nu_{\gamma}\neq0}\gamma$ of $G$.

In either formulation, we attain operational interpretations of the basic
results of DHR-DR theory, which provide the physical interpretation of any
state $\omega\in\mathfrak{f}(\pi)$ as a mixture of the DHR-selected states,
with respect to their internal-symmetry aspects, specifying its $G$%
\textit{\textbf{-charge contents}} understood as the $G$-representation
contents. Since the spacetime behaviours of quantum fields are expressed by
the observable net $\mathcal{O}\longmapsto\mathfrak{A}(\mathcal{O})$ and since
the internal symmetry aspects are described in the above machinery also
encoded in $\mathfrak{A}$, the role of the field algebra $\mathfrak{F}$ and
the internal symmetry group $G$ becomes now quite subsidiary, simply providing
comprehensible vocabulary based on the covariant objects under the symmetry
transformations. Thus, we have arrived at an physical and operational picture
for the sector theory showing the parallelism with the previous discussion of
the thermal interpretation based upon the \textit{c}$\rightarrow$\textit{q
}channel\textit{\ }$\mathcal{C}:\mathcal{A}\rightarrow C(B_{K})$. While, in
the latter case, the reference system to provide the vocabulary for the
interpretation is already known at the beginning, it is remarkable that the
corresponding one, $\mathfrak{Z}_{\pi}(\mathfrak{A})\simeq l^{\infty}(\hat
{G})$, in the DHR-DR theory naturally emerges from the basic ingredients of
the theory written in terms of the algebra $\mathfrak{A}$ of observables,
through the chain of equivalence starting from the DHR criterion:
[DHR-selected representations of $\mathfrak{A}$] $\Longleftrightarrow$
[Doplicher-Roberts category $\mathcal{T}$ ] $\Longleftrightarrow$ [$Rep_{G}$
and $G$] $\Longrightarrow$ [$\hat{G}=Spec(\mathfrak{Z}_{\pi}(\mathfrak{A}))$].

From the above observation that the ambiguity in the choice of a cross section
$\hat{G}\ni\gamma\longmapsto\rho_{\gamma}\in\lbrack\rho_{\gamma}]\subset
\Delta$ which picks up one $\rho_{\gamma}$ to each $\gamma\in\hat{G}$ among
the equivalence class $\{Ad(u)\circ\rho_{\gamma};u\in\mathfrak{A}$:
unitary$\}$\ is essentially due to observables in $\mathfrak{A}$ related to
the spacetime symmetry, i.e., the energy contents of sectors, we realize that
it is important to understand the mutual relations between the energy-momentum
spectrum and the sectors as internal-symmetry spectrum, in such a form as the
\textit{\textbf{energy contents of sectors}}: for instance, the contents of
the sectors parametrized by $\gamma\in\hat{G}$, $\gamma\neq\iota$(: the
trivial representation corresponding to the vacuum sector) are \textit{excited
states }above the vacuum. Since only the sector with trivial representation
$\iota\in\hat{G}$ contains the vacuum state with the \textit{minimum }energy
$0$ and since all other sectors consist of the excited states, the above
picture suggests the following results to be expected to hold (under the
assumption of the existence of a mass gap):
\begin{equation}
\min\{Spec(\hat{P}_{0}\upharpoonright_{\mathfrak{H}_{0}})\}=0,\text{
\ \ \ }\inf\{Spec(\hat{P}_{0}\upharpoonright_{\mathfrak{H}_{0}^{\perp}})\}>0.
\end{equation}
In the treatment of thermal functions in Sec.2, it is easily seen that, while
the entropy density $s(\beta)$ is not contained in the image set
$\mathcal{C}(\mathcal{T}_{x})$ due to the absence of such a quantum observable
$\hat{s}(x)\in\mathcal{T}_{x}$ that $\omega_{\beta}(\hat{s}(x))=s(\beta)$, it
can be \textit{approximated} by the thermal functions in $\mathcal{C}%
(\mathcal{T}_{x})$. In order to facilitate the above discussions of mutual
relations between spacetime and internal symmetries, it is important to have
those observables freely at hand which detect the $G$%
-charge\textit{\textbf{\ }}contents in $\mathfrak{Z}_{\pi}(\mathfrak{A}%
)=l^{\infty}(\hat{G})$, and, for this purpose, we need also here to consider
the problem as to how such observables can be supplied from the \textit{local}
observables belonging to $\mathfrak{A}(\mathcal{O})$, i.e., the approximation
of global order parameters by local order fields or central sequences. For
this purpose, the analyses of point-like fields and the rigorous method of
their operator-product expansions developed by Bostelmann in \cite{Bo} would
be quite useful in these contexts. All the above sort of considerations (with
the modifications of the DHR selection criterion necessitated by the possible
presence of the long range forces, such as of Buchholz-Fredenhagen type) will
be crucially relevant to the approach to the colour confinement problem, and,
especially the latter one (to find a suitably modified criterion) seems to be
quite a non-trivial issue there.

\section{SSB-vacua as continuous sectors with order parameter whose quantum
precursor is Goldstone mode}

\subsection{Dual net $\mathfrak{A}^{d}$ and unbroken remaining symmetry $H$}

To treat physically more interesting cases of \textit{spontaneous symmetry
breakdown (SSB)},\textbf{\ }we need to extend the original DR sector theory
where the internal symmetry is unbroken with unitary implementers as long as
the Haag duality $\mathfrak{A}^{d}(\mathcal{O}):=\pi_{0}(\mathfrak{A}%
(\mathcal{O}^{\prime}))^{\prime}=\pi_{0}(\mathfrak{A}(\mathcal{O}))$ (for
$\mathcal{O}\in\mathcal{K}$) holds to play the crucial roles. It can be shown
that this property is also a necessary condition for the field system with
normal statistics and with unbroken symmetry (see, \cite{DHR, DR90}). As
pointed out by Roberts \cite{Roberts74}, SSB does not take place without the
breakdown of the Haag duality.

In the previous case with unbroken symmetry, the superselection sectors are
parametrized by the \textit{discrete} variables belonging to the dual $\hat
{G}$ of a compact group $G$. In the situation with SSB, one anticipates
physically the appearance of \textit{continuous\ }macroscopic \textit{order
parameters}, as typically exemplified by the continuous directions of
magnetization in the ferromagnetism, which strongly suggests the appearance of
\textit{continuous\ superselection sectors},\textit{\ }parametrized by
macroscopic order parameters\textit{. }This will be shown actually to be the
case in the following.

For the sake of convenience, we change the notation adopted in Sec.3 in the
unbroken symmetry case, so that the observable algebra $\mathfrak{A}$ and the
symmetry group $G$ in Sec.3 are replaced, respectively, by the dual net
$\mathfrak{A}^{d}$ (of the genuine observable algebra $\mathfrak{A}$) and the
group $H$ of \textit{unbroken remaining symmetry} in the present context. To
begin with, the correspondence between physically relevant states $\omega$
around the vacuum $\omega_{0}$ and such an endomorphism $\rho$ as
$\omega=\omega_{0}\circ\rho$ can be maintained when all the ingredients here
are understood in relation to the dual net $\mathfrak{A}^{d}$ under the
natural assumption of \textit{essential duality }%
\begin{equation}
\mathfrak{A}^{dd}=\mathfrak{A}^{d}%
\end{equation}
which is equivalent to the local commutativity of the dual net and is valid
whenever some Wightman fields are underlying the theory \cite{BDLR92}. First,
in view of the relation $\pi_{0}(\mathfrak{A}^{d}(\mathcal{O}^{\prime
}))^{\prime\prime}=\pi_{0}(\mathfrak{A}(\mathcal{O}^{\prime}))^{\prime\prime}$
\cite{Roberts74}, the starting vacuum state and representation, $\omega_{0}$
and $(\pi_{0},\mathfrak{H}_{0})$, can safely be extended from $\mathfrak{A}$
to $\mathfrak{A}^{d}$ (meaning both the local nets and the global algebras).
Then the DHR selection criterion is understood for the states $\omega$ of
$\mathfrak{A}^{d}$, as $\pi_{\omega}\upharpoonright_{\mathfrak{A}%
^{d}(\mathcal{O}^{\prime})}=\pi_{0}\upharpoonright_{\mathfrak{A}%
^{d}(\mathcal{O}^{\prime})}$, and is equivalent to the existence of $\rho
\in\mathcal{T}\subset End(\mathfrak{A}^{d})$ such that $\pi_{\omega}=\pi
_{0}\circ\rho$. On the basis of these items, we can repeat the same procedure
of constructing the field algebra $\mathfrak{F}$ and the group $H$ of unbroken
symmetry according to the general method \cite{DR89, DR90}:
\begin{equation}
\mathfrak{F}=\mathfrak{A}^{d}\underset{\mathcal{O}_{d_{0}}^{H}}{\otimes
}\mathcal{O}_{d_{0}},\text{ \ \ \ }H=Gal(\mathfrak{F/A}^{d}).
\end{equation}

\subsection{Spontaneously broken symmetry}

Now we start to clarify the sector structure associated with a spontaneously
broken symmetry described by the Galois group $G:=Gal(\mathfrak{F}%
/\mathfrak{A})\supset H$. First we consider the irreducible $H$-covariant
vacuum representation $(\pi,U,\mathfrak{H})$ of the system $\mathfrak{F}%
\underset{\tau}{\curvearrowleft}H$, $\pi(\tau_{h}(F))=U(h)\pi(F)U(h)^{\ast}$
for $\forall F\in\mathfrak{F}$, $\forall h\in H$, containing the original
representation $(\pi_{0},\mathfrak{H}_{0})$ of $\mathfrak{A}$ and of
$\mathfrak{A}^{d}$ as the cyclic fixed-point subspace under $U(H)$:
$\mathfrak{H}_{0}=\{\xi\in\mathfrak{H};$ $U(h)\xi=\xi$ for $\forall h\in H\}$,
$\overline{\pi(\mathfrak{F)H}_{0}}=\mathfrak{H}$. Then\ according to the DHR
sector structure in the unbroken case \cite{DHR}, we have%
\begin{equation}
\mathfrak{Z}_{\pi}(\mathfrak{A}^{d})=\mathfrak{Z}(U(H)^{\prime\prime
})=\underset{\eta\in\hat{H}}{\oplus}\mathbb{C}(\mathbf{1}_{\mathfrak{H}_{\eta
}}\otimes\mathbf{1}_{W_{\eta}})=l^{\infty}(\hat{H}).
\end{equation}
Since this group $H$ is the maximal group of unbroken symmetry in the
irreducible vacuum situation, the group $G$ bigger than $H$ cannot be
unitarily implemented in the above representation $(\pi,\mathfrak{H})$ of
$\mathfrak{F}$, which is just the precise meaning of the SSB of $G$ in the
present situation. To cover more general situations we propose a general
definition of SSB in the following form:

\begin{definition}
\label{Def:SSB}A symmetry described by a (strongly continous) automorphic
action $\tau$ of $G$ on the field algebra $\mathfrak{F}$ is said to be
\textbf{unbroken }in a given representation $(\pi,\mathfrak{H})$ of
$\mathfrak{F}$ if the spectrum of the centre $\mathfrak{Z}_{\pi}%
(\mathfrak{F})=\mathfrak{Z}(\pi(\mathfrak{F)}^{\prime\prime})$ is pointwise
invariant under the action of $G$ induced on $Spec(\mathfrak{Z}_{\pi
}(\mathfrak{F}))$ (almost everywhere w.r.t. the central measure $\mu$ which
appears in the central decomposition of $\pi$ into factor representations). If
the symmetry is not unbroken in $(\pi,\mathfrak{H})$, it is said to be
\textbf{broken spontaneously} there.
\end{definition}

In particular, $G$ acting on $\mathfrak{F}$ is unbroken\textbf{\ }if each
\textbf{factor} subrepresentation $(\sigma,\mathfrak{H}_{\sigma})$,
$\sigma(\mathfrak{F)}^{\prime}\cap\sigma(\mathfrak{F)}^{\prime\prime
}=\mathbb{C}\mathbf{1}_{\mathfrak{H}_{\sigma}}$, appearing in the central
decomposition of $(\pi,\mathfrak{H})$ admits a \textbf{covariant
representation} of the system $G\overset{\tau}{\curvearrowright}\mathfrak{F}$
in terms of a (strongly continuous) unitary representation $(U_{\sigma
},\mathfrak{H}_{\sigma})$ of $G$ verifying the relation $\sigma(\tau
_{g}(F))=U_{\sigma}(g)\sigma(F)U_{\sigma}(g)^{\ast}$ for $\forall g\in
G,\forall F\in\mathfrak{F}$.

\begin{remark}
In essence, SSB\ means the \textit{conflict between unitary implementability
and factoriality (=triviality of centres)} \cite{IO99}. The situation with SSB
is seen to exhibit the features of the so-called \textit{\textquotedblleft
infrared instability\textquotedblright\textbf{\ }}under the action of $G$,
because $G$ does not stabilize the spectrum of centre which can be viewed
physically as \textit{macroscopic order parameters }emerging in the infrared
(=low energy) regions.
\end{remark}

\begin{remark}
Since the above definition of SSB still allows the mixture of unbroken and
broken \textrm{sub}representations of a given $\pi$, we need to decompose
$Spec(\mathfrak{Z}_{\pi}(\mathfrak{F}))$ into $G$-invariant domains which
cannot be further decomposed. It is easily seen that each such
\textrm{minimal} domain is characterized by the \textit{ergodicity} under $G$
which is nothing but the notion of \textrm{central ergodicity}. Then $\pi$ is
decomposed into the direct sum (or, direct integral) of \textit{unbroken
factor representations} and \textit{broken non-factor representations}, each
component of which is stable under $G$. In this way we obtain a \textrm{phase
diagram}\textit{ }on the spectrum of the centre.
\end{remark}

As indicated above, the natural physical picture of \textit{order parameters
arising from the SSB} from $G$ down to $H$ is realized in connection with the
sector structure of the whole theory involving the presence of
\textit{continuous sectors} parametrized by $\dot{g}:=Hg\in H\backslash G$.
Here we need to combine the above two formulations of discrete sectors of
unbroken internal symmetry (Sec.3) and of continuous sectors (Sec.2) in the
following way. One important point to be mentioned is that our motivation for
treating here the \textit{centre}s at various levels of representations is
always coming from the natural and inevitable occurrence of \textit{disjoint}
representations which leads to the appearance of \textit{macroscopic order
parameters to classify different modes of }macroscopic manifestations of
microscopic systems; this should be properly contrasted to a mathematical
pursuit of generalizing the pre-existing machinery involving factor algebras
to non-factorial ones.

According to this formulation, we should find such a covariant representation
of the system $(\mathfrak{F}\underset{\tau}{\curvearrowleft}G)$ as
implementing \textit{minimally} the broken $G$ in the sense of \textit{central
ergodicity }under $G$. Since the subgroup $H$ is unbroken in the irreducible
covariant representation $(\pi,U,\mathfrak{H)}$ of $\mathfrak{F}\underset
{\tau}{\curvearrowleft}H$, what we seek for can actually be provided by the
representation $(\hat{\pi},\mathfrak{\hat{H}})$, induced from $(\pi
,U,\mathfrak{H)}$, of the crossed product $\mathfrak{\hat{F}}:=\mathfrak{F}%
\rtimes(H\backslash G)=\Gamma(G\times_{H}\mathfrak{F})$ of $\mathfrak{F}$ with
the homogeneous space $H\backslash G$ (having the right $G$-action being
transitive, and hence, trivially $G$-ergodic), which can be identified with
the algebra of $H$-equivariant norm-continuous functions $\hat{F}%
:G\rightarrow\mathfrak{F}$ satisfying
\begin{equation}
\hat{F}(hg)=\tau_{h}(\hat{F}(g)). \label{equivA}%
\end{equation}
The action $\hat{\tau}$ of $G$ on $\hat{F}\in\mathfrak{\hat{F}}$ is defined
by
\begin{equation}
\lbrack\hat{\tau}_{g}(\hat{F})](g_{1})=\hat{F}(g_{1}g),
\end{equation}
with $g,g_{1}\in G$, consistently with (\ref{equivA}). For the technical
reason, we need here the assumption that $G$ should be a \textit{locally
compact} group equipped with a left-invariant Haar measure, although the
general definition as a Galois group $G:=Gal(\mathfrak{F}/\mathfrak{A})$ does
not ensure it. Then denoting $d\xi$ the left-invariant Haar measure on $G/H$
(equipped with the \textit{left }$G$-action), we define a Hilbert space
$\mathfrak{\hat{H}}$ of $L^{2}$-sections of $G\times_{H}\mathfrak{H}$ by%
\begin{equation}
\mathfrak{\hat{H}}=\int_{\xi\in G/H}^{\oplus}(d\xi)^{1/2}\mathfrak{H}%
=\Gamma_{L^{2}}(G\times_{H}\mathfrak{H},d\xi),
\end{equation}
which can also be identified with the $L^{2}$-space of $\mathfrak{H}$-valued
$(U,H)$-equivariant functions $\psi$ on $G$,
\begin{equation}
\psi(gh)=U(h^{-1})\psi(g)\text{ \ \ \ for }\psi\in\mathfrak{\hat{H}}\text{,
}g\in G\text{, }h\in H. \label{equivS}%
\end{equation}
On this $\mathfrak{\hat{H}}$, representations $\hat{\pi}$ and $\hat{U}$ of
$\mathfrak{\hat{F}}$ and $G$ are defined, respectively, by
\begin{align}
(\hat{\pi}(\hat{F})\psi)(g)  &  :=\pi(\hat{F}(g^{-1}))(\psi(g))\text{
\ \ \ for }\hat{F}\in\mathfrak{\hat{F}}\text{, }\psi\in\mathfrak{\hat{H}%
}\text{, }g\in G,\label{crss-rep}\\
(\hat{U}(g_{1})\psi)(g)  &  :=\psi(g_{1}^{-1}g)\text{ \ \ \ for }g,g_{1}\in G,
\end{align}
which are compatible with the above equivariance condition (\ref{equivS}):
\begin{align}
&  (\hat{\pi}(\hat{F})\psi)(gh)=\pi(\hat{F}(h^{-1}g^{-1}))(\psi
(gh))\nonumber\\
&  =U(h^{-1})\pi(\hat{F}(g^{-1}))U(h)U(h^{-1})(\psi(g))=U(h^{-1})(\hat{\pi
}(\hat{F})\psi)(gh);\\
&  (\hat{U}(g_{1})\psi)(gh)=\psi(g_{1}^{-1}gh)=U(h^{-1})\psi(g_{1}%
^{-1}g)=U(h^{-1})(\hat{U}(g_{1})\psi)(g),
\end{align}
and satisfies the covariance relation:
\begin{equation}
\hat{\pi}(\hat{\tau}_{g}(\hat{F}))=\hat{U}(g)\hat{\pi}(\hat{F})\hat{U}%
(g)^{-1}.
\end{equation}

We consider an embedding $\hat{\imath}_{H\backslash G}:\mathfrak{F}%
\hookrightarrow\mathfrak{\hat{F}}$ of $\mathfrak{F}$ into $\mathfrak{\hat{F}}$
defined by
\begin{equation}
\lbrack\hat{\imath}_{H\backslash G}(F)](g):=\tau_{g}(F),
\end{equation}
which intertwines the $G$-actions $\tau$ on $\mathfrak{F}$ and $\hat{\tau}$ on
$\mathfrak{\hat{F}}$,
\begin{equation}
\hat{\imath}_{H\backslash G}\circ\tau_{g}=\hat{\tau}_{g}\circ\hat{\imath
}_{H\backslash G}\text{\ \ \ }(\forall g\in G).
\end{equation}
Combining $\hat{\imath}_{H\backslash G}$ with $\hat{\pi}$, we obtain a
covariant representation $(\bar{\pi},\hat{U},\mathfrak{\hat{H}})$, $\bar{\pi
}:=\hat{\pi}\circ\hat{\imath}_{H\backslash G}$, of $\mathfrak{F}\underset
{\tau}{\curvearrowleft}G$ defined on $\mathfrak{\hat{H}}$ by
\[
(\bar{\pi}(F)\psi)(g):=\pi(\tau_{g^{-1}}(F))\psi(g)\text{ \ \ \ \ }%
(F\in\mathfrak{F},\psi\in\mathfrak{\hat{H}})
\]
and satisfying
\[
\bar{\pi}(\tau_{g}(F))=\hat{U}(g)\bar{\pi}(F)\hat{U}(g)^{-1}.
\]

All the above operations are compatible with the constraints of $H$-equivariance.

\subsection{Sector structures and \textit{c}$\rightarrow$\textit{q} channel}

The crucial information for determining the sector structure is the
\textit{centres} of $\mathfrak{A}^{d},\mathfrak{A}$ and $\mathfrak{F}$ in the
representation $(\bar{\pi},\mathfrak{\hat{H}})$. The mutual relations among
the relevant C*-algebras can be summarized in the following commuting diagram:%

\[%
\begin{array}
[c]{ccc}
& \mathfrak{\hat{F}}=\Gamma(G\underset{H}{\times}\mathfrak{F}) & \\
^{\hat{\imath}_{H\backslash G}}\nearrow &  & _{\hat{\imath}_{H\backslash G}%
}\nwarrow\\
\mathfrak{A}^{d}=\mathfrak{F}^{H} & \overset{m_{H}}{\underset{i_{H}%
}{\leftrightarrows}} & \mathfrak{F}=\mathfrak{A}^{d}\underset{\mathcal{O}%
_{d_{0}}^{H}}{\otimes}\mathcal{O}_{d_{0}}\\
_{m_{G/H}}\searrow\nwarrow^{i_{G/H}} &  & ^{i_{G}}\nearrow\swarrow_{m_{G}}\\
& \mathfrak{A}\subset\mathfrak{F}^{G} &
\end{array}
,
\]
where the maps $i_{G}$ and $m_{G}$, etc. are, respectively, the embedding maps
(of a C*-algebra into another) and the conditional expectations, such as
\begin{equation}
m_{G/H}:\mathfrak{A}^{d}=\mathfrak{F}^{H}\ni B\longmapsto m_{G/H}%
(B):=\int_{G/H}d\dot{g}\ \tau_{g}(B)\in\mathfrak{F}^{G}\mathfrak{.}%
\end{equation}
Using the relations, $\bar{\pi}(\mathfrak{A})=\int_{\dot{g}\in H\backslash
G}^{\oplus}d\dot{g}$ $\pi(\mathfrak{A})=\mathbf{1}_{L^{2}(G/H,d\dot{g}%
)}\otimes\pi(\mathfrak{A})$ and $\pi(\mathfrak{A})^{\prime\prime}%
=\pi(\mathfrak{A}^{d})^{\prime\prime}$ (following from $\pi(\mathfrak{A}%
^{d})=\pi(\mathfrak{A})^{\prime\prime}\cap\pi(\mathfrak{F})$ \cite{BDLR92}),
we obtain

\begin{proposition}
\cite{Oji2002}%
\begin{align*}
\mathfrak{Z}_{\bar{\pi}}(\mathfrak{F})  &  =L^{\infty}(H\backslash G;d\dot
{g})=\mathfrak{Z}_{\hat{\pi}}(\mathfrak{\hat{F}});\\
\mathfrak{Z}_{\bar{\pi}}(\mathfrak{A})  &  =\mathbf{1}_{L^{2}(G/H,d\dot{g}%
)}\otimes\mathfrak{Z}_{\pi}(\mathfrak{A})=\mathbf{1}_{L^{2}(G/H,d\dot{g}%
)}\otimes l^{\infty}(\hat{H});\\
\mathfrak{Z}_{\bar{\pi}}(\mathfrak{A}^{d})  &  =L^{\infty}(H\backslash
G;d\dot{g})\otimes\mathfrak{Z}_{\pi}(\mathfrak{A}^{d})=L^{\infty}(H\backslash
G;d\dot{g})\otimes l^{\infty}(\hat{H}).
\end{align*}

\end{proposition}

(The first line follows from the disjointness $\pi\overset{\shortmid}{\circ}$
$(\pi\circ\tau_{g})$ (for $\forall g\in G\diagdown H$) and the definition of
$\bar{\pi}=\hat{\pi}\circ\hat{\imath}_{H\backslash G}$, and hence,
$\mathfrak{Z}_{\bar{\pi}}(\mathfrak{F})=L^{\infty}(H\backslash G;d\dot
{g})\subset\bar{\pi}(\mathfrak{F})^{\prime\prime}\cap\hat{U}(H)^{\prime}%
=\bar{\pi}(\mathfrak{A}^{d})^{\prime\prime}$ from which the third line follows.)

\begin{remark}
It is remarkable that only the centre of von Neumann algebra $\bar{\pi
}(\mathfrak{A}^{d})^{\prime\prime}$ representing $\mathfrak{A}^{d}$ in
$(\bar{\pi},\mathfrak{\hat{H}})$ carries full information on \textbf{both}%
\textit{\ aspects} of order parameters of broken $H\backslash G$ and of
unbroken $\hat{H}$, whereas centres of any other von Neumann algebras
representing $\mathfrak{\hat{F}}$, $\mathfrak{F}$ or $\mathfrak{A}$ in
$(\pi,\mathfrak{H})$ or $(\bar{\pi},\mathfrak{\hat{H}})$ carry \textit{only
partial information}.
\end{remark}

On the basis of these structures of relevant centres of representations, we
define a \textit{c}$\rightarrow$\textit{q} \textit{channel} $\Psi$ as
follows:
\begin{align}
&  \Psi:\mathfrak{A}^{d}\ni B\longmapsto\Psi(B)\in C(H\backslash
G)\otimes\mathfrak{Z}_{\pi}(\mathfrak{A}^{d}),\nonumber\\
&  [\Psi(B)](\dot{g},\eta):=(\omega_{0}\circ\rho_{\eta}\circ m_{H}%
)(\tau_{g^{-1}}(B))\text{\ }\nonumber\\
&  \text{ \ \ \ \ \ \ \ \ \ \ \ \ \ \ for }(\dot{g},\eta)\in(H\backslash
G)\times Spec(\mathfrak{Z}_{\pi}(\mathfrak{A}^{d})).
\end{align}
Here, $\rho_{\eta}\in\mathcal{T}$ is a local endomorphism of $\mathfrak{A}%
^{d}$ belonging to the DR-category $\mathcal{T}_{\mathfrak{A}^{d}}%
\mathcal{\ }$on $\mathfrak{A}^{d}$ and $g\in G$ is an arbitrary representative
of $\dot{g}=Hg\in H\backslash G$.

\begin{remark}
It is due to the \textbf{non-invariance} of $\pi$ under $G$ that the validity
of such relations as $\pi(\mathfrak{A})^{\prime\prime}=\pi(\mathfrak{A}%
^{d})^{\prime\prime},\mathfrak{Z}_{\pi}(\mathfrak{A})=\mathfrak{Z}_{\pi
}(\mathfrak{A}^{d})=l^{\infty}(\hat{H})$ is consistent with the $G$-invariance
of $\mathfrak{A}(\subset\mathfrak{F}^{G})$ and the non-invariance of
$\mathfrak{A}^{d}=\mathfrak{F}^{H}$.
\end{remark}

Before a field algebra $\mathfrak{F}$ is constructed, the Doplicher-Roberts
method based on the local endomorphisms and the related Cuntz algebras
\cite{Cuntz} seems to be the\ only possible path starting from $\mathfrak{A}$
to the pair of $\mathfrak{F}$ and $G$ \textit{without knowing} either of them,
which has necessarily led us to an \textit{unbroken }and \textit{compact
}symmetry group. However, in the situation of SSB with possible presence of
massless spectrum, there is \textit{no} reason \textit{nor }guarantee for the
broken group $G=Gal(\mathfrak{F}/\mathfrak{A})$ to be compact, as shown in
\cite{BDLR92} through the counter-examples. Fortunately, once $\mathfrak{F}$
is so constructed from the dual net $\mathfrak{A}^{d}$ and the DR category
$\mathcal{T}_{\mathfrak{A}^{d}}$ as to show certain kinds of stability
properties (as will be discussed later), we need not stick any more to the
original line of thought inherent to the Doplicher-Roberts theory: having at
hand the information on the group $G=Gal(\mathfrak{F}/\mathfrak{A})$, we can
control the mutual relations among $\mathfrak{F}$, $G$, $\mathfrak{A}^{d}$ and
$\mathfrak{A}$ by means of various versions of crossed products applicable to
$G$, \textit{irrespective} of whether it is compact or not \cite{Nak-Take} (as
long as it is assumed to be locally compact).

When the group $G=Gal(\mathfrak{F}/\mathfrak{A})$ of spontaneously broken
symmetry is compact, as is common in the physical examples of SSB (such as the
case of chiral $SU(2)\times SU(2)$ down to the vectorial $SU(2)$), we can get
more information. We see here the important roles played by the
\textit{finite-dimensional induction/reduction} between $H$ and $G$ on the
algebra $\mathfrak{F}$ according to the following results under the assumption
of the compactness of $G$:

\begin{enumerate}
\item \textit{Finite-dimensional induction} for a compact pair $H$
$\hookrightarrow G$: \newline Any finite-dimensional unitary representation
$(\eta,W)$ of $H$ can be extended to a representation $(\gamma,V)$ of $G$ by
taking a direct sum $\gamma|_{H}\cong\eta\oplus\eta^{\prime}$ with a suitable
representation $(\eta^{\prime},W^{\prime})$ of $H$ (for proof, see
\cite{TodaMimura}). At the level of a field algebra, this kind of induction is
sufficient, in contrast to the situations of states for which the genuine
Mackey induction involving infinite dimensional spaces is indispensable.

\item Stability and consistency of field algebra construction in SSB cases:
\newline In use of the above result, one can verify the \textit{stability} of
the crossed product construction of the field algebra under the change of
Cuntz algebras.

\begin{proposition}
\cite{NO} If the dual-net algebra $\mathfrak{A}^{d}$ is properly infinite
C*-algebra, the field algebra $\mathfrak{F}$ due to the original DR
construction from $\mathfrak{A}^{d}$ and a Cuntz algebra $\mathcal{O}_{d_{0}}$
is isomorphic to the crossed product of $\mathfrak{A}^{d}$ with a Cuntz
algebra $\mathcal{O}_{d}$ for any $d>d_{0}$:
\begin{equation}
\mathfrak{F}:=\mathfrak{A}^{d}\underset{\mathcal{O}_{d_{0}}^{H}}{\otimes
}\mathcal{O}_{d_{0}}\cong\mathfrak{A}^{d}\underset{\mathcal{O}_{d}^{H}%
}{\otimes}\mathcal{O}_{d}.
\end{equation}

\end{proposition}

With this freedom, we can naturally let $\mathfrak{F}$ be acted upon by a
compact Lie group $G$ whose fundamental representation is $d$-dimensional,
bigger than the corresponding dimensionality $d_{0}$ of the unbroken $H$. In
this situation, while the relation $g(\mathfrak{A}^{d})=\mathfrak{A}%
^{d}=\mathfrak{F}^{\tau(H)}$ for $g\in G$ forces $g$ to belong to the
normalizer $N_{H}=\{s\in G;sHs^{-1}\subset H\}$ of unbroken $H$ in $G$, the
equality
\begin{equation}
g(\mathfrak{A}^{d})\underset{\mathcal{O}_{d}^{gHg^{-1}}}{\otimes}%
g(\mathcal{O}_{d})=g(\mathfrak{A}^{d}\underset{\mathcal{O}_{d}^{H}}{\otimes
}\mathcal{O}_{d})=\mathfrak{F}%
\end{equation}
can be verified \cite{NO} even for such $g\in G$ that $g\notin$ $N_{H}$, which
shows the \textit{consistency} of the construction method with the action of
$G$ bigger than $H$. \newline(In \cite{BDLR93} where the relation
$Gal(\mathfrak{A}^{d}/\mathfrak{A})=N_{H}/H$ was verified, the analysis of
degenerate vacua was restricted to $N_{H}$ in order to avoid $g(\mathfrak{A}%
^{d})\neq\mathfrak{A}^{d}$. In the physically interesting situations involving
\textit{Lie groups}, however, the reductivity of a compact Lie group $H$
implies that $N_{H}/H$ is abelian and/or discrete with vanishing Lie brackets,
which does not seem to be relevant to the physically meaningful contexts.)

\item \textit{Duality for homogeneous spaces}: corresponding to the relevance
of the Tannaka-Krein duality between a compact group and the category of its
representations in the DR construction, we encounter here its extended version
to a homogeneous space $H\backslash G$. \newline\ \ \ \ \ For a compact group
pair $H\hookrightarrow G$, the definition of $Rep_{H\backslash G}$ and the
mutual relations among $Rep_{G}$, $Rep_{H}$ and $Rep_{H\backslash G}$ can be
described in terms of a \textit{homotopy-fibre category} $Rep_{G}$ over
$Rep_{H}$ with $Rep_{H\backslash G}$ as homotopy fibre (\cite{Maum}): over
$\eta\in Rep_{H}$ a homotopy fibre (h-fibre for short) is given by a category
$\eta/Rep_{G}$ (which is called a comma category under $\eta$ \cite{MacL}
whose objects are pairs $(\gamma,T)$ of $\gamma\in$ $Rep_{G}$ and $T\in
Rep_{H}(\eta,\gamma|_{H})$ and whose morphisms $\phi:(\gamma,T)\rightarrow
(\gamma^{\prime},T^{\prime})$ are given by $\phi\in Rep_{G}(\gamma
,\gamma^{\prime})$ s.t. $T^{\prime}=\phi\circ T$:
\begin{equation}%
\begin{array}
[c]{ccccc}
&  & \eta &  & \\
& \swarrow_{T} &  & _{T^{\prime}}\searrow & \\
\gamma|_{H} & \rightarrow & \underset{\phi|_{H}}{\rightarrow} & \rightarrow &
\gamma^{\prime}|_{H}\\
r_{H}\uparrow\text{\ \ } &  &  &  & \text{ \ }\uparrow r_{H}\\
\gamma & \rightarrow & \underset{\phi}{\rightarrow} & \rightarrow &
\gamma^{\prime}%
\end{array}
\end{equation}
(To be more precise, the comma category $\eta/Rep_{G}$ is to be understood as
$\eta/r_{H}$ where the functor $r_{H}:Rep_{G}\rightarrow Rep_{H}$ is the
restriction of $G$-representations to the subgroup $H$ of $G$.)\newline%
\ \ \ \ The h-fibre over the trivial representation $\eta=\iota\in Rep_{H}$ of
$H$ is nothing but the category of \textit{linear representations of
}$H\backslash G$ due to Iwahori-Sugiura \cite{IwaSug}, to which any other
h-fibres can be shown to be homotopically equivalent \cite{Maum}.

\item Generalizing a theorem in \cite{DR89}, we obtain a result to construct
the extended field algebra $\mathfrak{\hat{F}}$ for implementing the broken
$G$ also as a crossed product with a Cuntz algebra:

\begin{proposition}
\cite{Oji2002} If $\mathfrak{Z}(\mathfrak{A}^{d})=\mathbb{C}\mathbf{1}$ (as a
C*-algebra), we have the following relations
\begin{align}
&  \mathfrak{A}^{d}\underset{\mathcal{O}_{d}^{G}}{\otimes}\mathcal{O}%
_{d}=\Gamma(G\underset{H}{\times}(\mathfrak{A}^{d}\underset{\mathcal{O}%
_{d}^{H}}{\otimes}\mathcal{O}_{d}))=\mathfrak{\hat{F}},\\
&  \mathfrak{\hat{F}}^{\hat{\tau}(G)}=\mathfrak{A}^{d}\underset{\mathcal{O}%
_{d}^{G}}{\otimes}\mathbf{1},\\
&  Spec(\mathfrak{Z}(\mathfrak{A}^{d}\underset{\mathcal{O}_{d}^{G}}{\otimes
}\mathcal{O}_{d}))=H\backslash G.
\end{align}

\end{proposition}
\end{enumerate}

The original version \cite{DR89} of the above relation was formulated for
$G=SU(d)$ and was used to detect the unbroken $H$ as the stabilizer of the
factorial subrepresentations. As for the physical significance of the extended
algebra $\mathfrak{\hat{F}}$ see the next subsection.

\subsection{Interpretation of sector structure: degenerate vacua with order
parameters, Goldstone modes and condensates}

Combining the previous two cases with purely continuous sectors (Sec.2) and
with purely discrete sectors (Sec.3), we can now adapt the present scheme for
treating generalized sectors to the situation with SSB in order to clarify the
sector structure involved there and the physical meaning of each ingredient
appearing so far in our fomulation of SSB.

The map
\[
\Psi^{\ast}:M_{1}(H\backslash G)\otimes M_{1}(\hat{H})\rightarrow
E_{\mathfrak{A}^{d}},
\]
defined as the dual of $\ \Psi:\mathfrak{A}^{d}\ni B\longmapsto\Psi(B)$,
$[\Psi(B)](\dot{g},\eta)=(\omega_{0}\circ m_{H\backslash G}\circ\rho_{\eta
}\circ m_{H})(\tau_{g^{-1}}(B))$, gives a \textit{c}$\rightarrow$\textit{q
channel}, whose inverse $(\Psi^{\ast})^{-1}$ exists on the mixtures
$\in\mathfrak{f}(\bar{\pi})$ of states on $\mathfrak{A}^{d}$ selected by the
DHR criterion, as a \textit{q}$\rightarrow$\textit{c channel} to provide the
physical interpretations of such states in terms of the order parameters in
$\dot{g}\in H\backslash G$, $H$-charge $\eta\in\hat{H}$. In view of our
starting premise of the observable algebra $\mathfrak{A}$, however, it looks
natural to take $A\in\mathfrak{A}$ as the argument of $\Psi$, instead of
$B\in\mathfrak{A}^{d}$. Because of $G$-invariance of $A\in\mathfrak{A}%
\subset\mathfrak{F}^{G}$, however, $\Psi\upharpoonright_{\mathfrak{A}}$ is
independent of $\dot{g}\in H\backslash G$, $[\Psi(A)](\dot{g},\eta
)=(\omega_{0}\circ m_{H\backslash G}\circ\rho_{\eta}\circ m_{H})(\tau_{g^{-1}%
}(A))=(\omega_{0}\circ m_{H\backslash G}\circ\rho_{\eta})(A)$, failing to pick
up the information on $H\backslash G$, so we here take $\mathfrak{A}^{d}$ as
our \textit{extended observables}. This standpoint is justified by the natural
physical meaning of $\mathfrak{A}^{d}$ as the \textit{maximal local net
generated by the original net }$\mathfrak{A}$, which is just a version,
adapted to the observable net, of the notion of the Borchers classes
\cite{Borch60} consisting of all the \textit{relatively local} fields to
absorb the \textit{arbitrariness in the \textquotedblleft interpolating
fields\textquotedblright\ }\cite{Nishi}. What we see here is that, in spite of
their $G$-noninvariance property, the \textit{Goldstone modes} related to the
homogeneous space $H\backslash G$ are allowed to appear here with the
qualification as such extended observables belonging to $\mathfrak{A}^{d}$ and
they detect the information concerning the position of a pure vacuum $\dot
{g}\in H\backslash G$ among the degenerate vacua, as is exhibited through the
$\dot{g}$-dependence of $\Psi(B)$ for $B\in\mathfrak{A}^{d}$.

To understand the situation, we first consider the physical meaning of the
obtained structures related to order parameters:

\begin{itemize}
\item[i)] $H\backslash G$ as \textit{order parameters} to parametrize the
degenerate vacua: in the decomposition of the representation space
$\mathfrak{\hat{H}}$ of $\mathfrak{\hat{F}}$ to pure vacuum representations of
$\mathfrak{F}$ in $\mathfrak{H}$, we get the centre $L^{\infty}(H\backslash
G,d\dot{g})=\mathfrak{Z}_{\bar{\pi}}(\mathfrak{F})=\mathfrak{Z}_{\hat{\pi}%
}(\mathfrak{\hat{F}})$ with the spectrum $H\backslash G$ which parametrizes
the \textbf{degenerate vacua}\textit{\ }with minimum energy $0$ generated by
the SSB of $G$ up to the unbroken remaining $H$. Mathematically, the Mackey
induction from $H$ to $G$ is relevant here. The physical meaning of
$H\backslash G$ is seen through such examples as the directions of
magnetization in the Heisenberg ferromagnets, or, as the \textit{Josephson
effect} where the difference of the phases of Cooper pair condensates between
adjacent vacua across a junction exhibits such eminent physical effects as the
resistance-free \textit{Josephson current}\textbf{\ }(see, e.g.,
\cite{Oji98}). In the context of \textit{static} structures of sectors, the
field algebra $\mathfrak{\hat{F}}$ bigger than $\mathfrak{F}$ looks redundant,
whereas it becomes relevant in the situations with the \textit{order
parameters behaving dynamically} as in the above cases through the couplings
with external fields (see, e.g., \cite{Froeh92}). Also the non-trivial centre
$C(H\backslash G)$ in $\mathfrak{\hat{F}}$ \textit{in the C*-version} with a
\textit{continuous} $G$-action resolves a puzzling conflict between the
disjointness $\omega\overset{\shortmid}{\circ}(\omega\circ\tau_{g})$ (for
\textit{any }$g\in G$ s.t. $g\notin H$) along the $G$-orbit of any pure vacuum
$\omega$ of $\mathfrak{F}$ and the continuous behaviours of the order
parameter $\dot{g}\in H\backslash G$ under $G$.

\item[ii)] Internal spectrum $\hat{H}$ of \textit{excited states} on a chosen
pure vacuum specified by a fixed $\dot{g}=Hg\in H\backslash G$: in the
representation space $\mathfrak{H}$ of $\mathfrak{F}$, we see the standard
picture of sectors $(\pi_{\eta},\mathfrak{H}_{\eta})$ with respect to
$\mathfrak{A}^{d}$ parametrized by $\eta\in\hat{H}$, which describes the
internal symmetry aspects of excited states in terms of the unbroken $H$ (to
be precise, $g^{-1}Hg\simeq H$ at $\dot{g}$) just in the same way as the
situations discussed in Sec. 3. For the description of this aspect, we find no
essential difference between $\mathfrak{A}$ and $\mathfrak{A}^{d}$, because of
the relations $\pi(\mathfrak{A})^{\prime\prime}=\pi(\mathfrak{A}^{d}%
)^{\prime\prime}$, $\mathfrak{Z}_{\pi}(\mathfrak{A})=\mathfrak{Z}_{\pi
}(\mathfrak{A}^{d})=l^{\infty}(\hat{H})$, valid in the factor representation
$(\pi,\mathfrak{H})$ of $\mathfrak{F}$ carrying no explicit information on the
order parameters i) of SSB.

\item[iii)] \textit{Goldstone modes} responsible for the gap $\mathfrak{A}%
(\mathcal{O})\subsetneqq\mathfrak{A}^{d}(\mathcal{O})$ at the \textit{local}
level, whose \textit{global} manifestation is found in $\bar{\pi}%
(\mathfrak{A})^{\prime\prime}\subsetneqq\bar{\pi}(\mathfrak{A}^{d}%
)^{\prime\prime}$ (both involving $H\backslash G$): in sharp contrast with the
above ii), we are concerned here with the algebraic dual objects of the
physically relevant order parameters of SSB appearing in i). The origin of
these gaps can be understood naturally in the following context: for a
co-action $\delta$ of $\hat{G}$ on $\mathfrak{F}^{G}$ the relations
\begin{align}
\mathfrak{F}  &  =\mathfrak{F}^{G}\rtimes_{\delta}\hat{G},\text{ }\nonumber\\
\mathfrak{A}^{d}  &  =\mathfrak{F}^{H}=\mathfrak{F}^{G}\rtimes_{\delta
}\widehat{(H\backslash G)},\text{\ } \label{Goldstn}%
\end{align}
are known to hold in the W*-version of crossed products \cite{Nak-Take}. If
its C*-version is verified, the latter relation (\ref{Goldstn}) shows that the
gap between $\mathfrak{A(\subset F}^{G}\mathfrak{)}$ and $\mathfrak{A}^{d}$
comes from the $G$-non-invariant elements in $\mathfrak{A}^{d}$ related to
$H\backslash G$, which can be interpreted properly as an abstract algebraic
form of the (would-be) \textit{Goldstone modes} (whose full-fledged form as
the massless Goldstone spectrum can be absent in the representation Hilbert
space $\mathfrak{H}$ depending upon the decay rates of long-range
correlations, as shown in \cite{BDLR92}). To understand this, recall that the
standard picture of Goldstone modes is given by the physical degrees of
freedom $\varphi$ responsible for the non-invariance of a chosen pure vacuum
$\omega_{0}$ under the action of broken $G$, $\omega_{0}(\tau_{g}%
(\varphi))\neq$ $\omega_{0}(\varphi)$ ($g\in G\diagdown H$), which yields, as
in i), the orbit $\{\omega_{0}\circ\tau_{g}$ ; $g\in G\}$ constituting of the
\textquotedblleft degenerate vacua\textquotedblright\ parametrized by $G/H$ as
the spectrum of the centre of $\mathfrak{A}^{d}$: $\mathfrak{Z}_{\bar{\pi}%
}(\mathfrak{A}^{d})=L^{\infty}(G/H)\otimes\mathfrak{Z}_{\pi}(\mathfrak{A}%
)=L^{\infty}(G/H)\vee\mathfrak{Z}_{\bar{\pi}}(\mathfrak{A})$. While neither
$\mathfrak{A}^{d}$ nor its local subalgebras contain non-trivial central
elements, there should exist some sequences of local elements (central
sequences\ or order fields) in $\mathfrak{A}^{d}$ tending to global central
elements belonging to $L^{\infty}(G/H)\subset\mathfrak{Z}_{\bar{\pi}%
}(\mathfrak{A}^{d})$, which is to be identified with the Goldstone modes
describing virtual transitions\ from one specific vacuum to another among
degenerate vacua. In view of the transformation property under $G$, it is
clear that this kind of sequences cannot be supplied by $\mathfrak{F}^{G}%
$($\supset\mathfrak{A}$), but should be found in the second component of
$\mathfrak{A}^{d}=\mathfrak{F}^{G}\rtimes\widehat{(H\backslash G)}$. Thus, the
main cause of the gap between $\mathfrak{A}^{d}$ and $\mathfrak{A}$ can be
found in the presence of the above sequences identified with Goldstone modes.
Then the relation (\ref{Goldstn}) can be interpreted as an algebraic version
of the Goldstone and/or low-energy theorems in the sense that they give a
\textit{dual} description of the SSB-sector structure with degenerate vacua in
i) in a local and/or algebraic virtual form (appearing already in a pure
vacuum); this will fully justify such a heuristic and physical expression that
\textquotedblleft Goldstone degrees of freedom related to $H\backslash G$
search the degenerate vacua in a virtual way\textquotedblright.
\end{itemize}

In the standard approach focusing on discrete sectors in ii), the continuous
sectors appearing in i) fail to be recognized as genuine sectors, as a
consequence of which the situations with continuous sectors only are regarded
as the \textit{absence of sectors }(cf. \cite{BDLR92}). From the above
discussion, however, we find both physical and mathematical reasons for
treating them as sectors, in view of the physically important roles played by
the associated order parameters as seen in i) and also of the mathematically
interesting interrelationship between i) and iii). It may be also instructive
to compare the above i) and ii) with the results in \cite{BDLR93}; analysis
was restricted there to the factor representation $(\pi,\mathfrak{H})$ of
$\mathfrak{F}$ without touching on the larger one $(\bar{\pi},\mathfrak{\hat
{H}})$ implementing the broken $G$, and hence, what is found as the degenerate
vacua is only along $N_{H}/H$ (with vanishing Lie algebra), failing to find
the whole $H\backslash G$-orbit. To recover the full information on the
degenerate vacua, one should not avoid the complications due to the
instability, $g(\mathfrak{A}^{d})\neq\mathfrak{A}^{d}$, of the dual-net
algebra $\mathfrak{A}^{d}$ which moves around inside $\mathfrak{F}$ under the
action of $G(\supsetneqq N_{H})$.

\begin{remark}
Since our focus in the above iii) is just as to how Goldstone modes appear in
$\mathfrak{A}^{d}$ as extended observables in spite of their $G$%
-non-invariance, the question whether $\mathfrak{A}$ is Galois-closed or not,
$\mathfrak{A}\overset{\text{?}}{=}\mathfrak{F}^{G}$, is irrelevant, with the
relation $\mathfrak{A}\subset\mathfrak{F}^{G}$ following from
$G:=Gal(\mathfrak{F}/\mathfrak{A})$ being sufficient. In the problem of the
intrinsic characterization of the observable net $\mathfrak{A}$ itself,
however, the problems as to whether this property holds or not, and, as to how
it is ensured, are interesting questions to be examined.
\end{remark}

To avoid possible confusions on the various notions appearing in SSB, one
needs to be careful about the distinctions and mutual relations among the
following four levels involving Goldstone modes and order parameters:

\begin{itemize}
\item[1)] degenerate vacua as continuous sectors parametrized by the
\textit{order parameters} $\dot{g}\in H\backslash G$ which is a global notion,

\item[2)] \textit{Goldstone modes} belonging to $\mathfrak{A}^{d}$, whose
massless spectrum (if any) is responsible for the validity of Goldstone theorem,

\item[3)] \textit{Goldstone multiplet} belonging to $\mathfrak{F}$ and
consisting of Goldstone modes together with \textit{condensates} responsible
for the above 1); this field multiplet transforms under $G$ according to a
\textit{linear} representation, which is nothing but a \textquotedblleft%
\textit{linear representation of a homogeneous space}\textquotedblright%
\ according to the definition of \cite{IwaSug}. What is most confusing is the
mutual relation between the Goldstone modes and the condensates; in the
simplest example of SSB from $G=SO(3)$ to $H=SO(2)$ with $H\backslash G=S^{2}
$ (e.g., Heisenberg ferromagnet), a pure vacuum among degenerate vacua is
parametrized and geometrically depicted by a point $p\in S^{2}$, a condensate
by a radius from the centre of the unit ball to $p$, and the Goldstone modes
geometrically expressed by \textit{tangent vectors} at $p$ \textit{tangential
to} $S^{2}$ and \textit{orthogonal to the condensate}. The Goldstone multiplet
is an entity in $\mathfrak{F}$ which is behaving as a three-dimensional
covariant vector under $SO(3)$.

\item[4)] There is a useful physical notion called \textquotedblleft nonlinear
realization\textquotedblright\ of Goldstone bosons \cite{CallaColeWess},
expressing the above situation in a geometric way and serving as very
effective tools in the derivation of the so-called low energy theorems, such
as the soft-pion theorem, to describe the low energy scattering processes
involving Goldstone bosons associated with SSB. While its functional role is
very akin to our Goldstone modes in 2), it may not be so straightforward to
accommodate it literally into the present context, because of the nonlinear
transformation law exhibited in its transformation property under $G$. In the
attempt to incorporate the general essence of low energy theorems into the
present context, however, this notion is expected to play some useful roles.
\end{itemize}

\section{Selection criteria as categorical adjunctions and their operational
meanings}

Here we emphasize the important roles played by the categorical adjunctions
underlying our discussions so far, in achieving the systematic organizations
of various domains in physics: the essence of the three formulae
(\ref{adjunction1}), (\ref{adjunction2}) and (\ref{adjunction}) encountered in
Sec.2 and Sec.3 can be summarized as follows:%
\[%
\begin{tabular}
[c]{|c|c|c|}\hline
$X\text{ (}=q\text{): to be classified}$ & $q\leftarrow c$ & $A\text{
(}=c\text{): to classify}$\\\hline
$x\underset{X}{\equiv}G(a)$ & $\underset{F}{\overset{G}{\leftrightarrows}}$ &
$F(x)\underset{A}{\equiv}a$\\\hline
$\text{selection criterion}$ & $q\rightarrow c$ & $\text{interpretation}%
$\\\hline
\end{tabular}
\
\]
with $X$ a quantum domain of generic states to be characterized and
classified, $A$ a classical classifying space identified with the spectrum of
centre, $G$ the \textit{c}$\rightarrow$\textit{q channel} and with $F$ the
\textit{q}$\rightarrow$\textit{c channel} to provide the interpretation of $X$
in terms of the vocabulary in $A$. This scheme exhibits the essential meaning
and the pertinence of this notion to our discussion of selecting, classifying
and interpreting physically interesting classes of states.

These cases, however, share such special features that the relevant categories
are groupoids of equivalence relations with all arrows invertible and that the
mapping between quantum and classical domains are groupoid isomorphisms, and
hence, the essence of adjunctions in our context is found in such quantitative
form as above. Perhaps this is because the category consisting of states of
C*-algebras is a rather rigid one, allowing only few meaningful morphisms
among different objects, requiring strict equalities or equivalence relations.
As we have seen above, however, once the contents of imposed selection
criteria are paraphrased into different languages, such as thermal functions
in Sec.2, the category of DHR-selected representations $\pi_{\omega}$, the
DR-category of local endomorphisms $\rho$ and that of group representations
$\gamma_{\rho}$ in Sec.3, then the machinery stored in the category theory
starts to work. In such contexts, objects are not always states, and arrows
between objects (taking such forms as intertwiners among local endomorphisms
or among group representations) or functors between different categories need
not necessarily be invertible. In the next subsections we also find that what
to be selected need not always be states but can be channels as well.

So we should not to mistake these special features of our examples as the
universal essence of the adjunction, especially because what is important
about categorical notions is their flexibility allowing to look at the same
object in many different ways and to unify objects with different appearances
in one and the same notion. Although we do not use it systematically here, we
give, for convenience, the general definition of adjunction \cite{MacL},
\begin{equation}
X(x,G(a))\simeq A(F(x),a), \label{adjunction3}%
\end{equation}
which involves four levels of notions, objects, arrows, a pair of functors
$G:A\rightarrow X$, $F:X\rightarrow A$ and a pair of natural transformations
(= arrows between functors) $\eta$ and $\varepsilon$ between two functors,
$\eta:1_{X}\overset{\cdot}{\rightarrow}GF$,$\ \varepsilon:FG\overset{\cdot
}{\rightarrow}1_{A}$, in such a way that the relation $\simeq$ is specified by
$\varepsilon_{F(x)}\circ F(\eta_{x})=1_{F(x)},$ $G(\varepsilon_{a})\circ
\eta_{G(a)}=1_{G(a)}$. This is equivalent to a natural family of bijections
$\nu_{x,a}:X(x,G(a))\simeq A(F(x),a)$ where \textquotedblleft
naturality\textquotedblright\ is characterized by the relations $\nu
_{y,b}(fgF(\psi))=G(f)\nu_{x,a}(g)\psi$ for $\forall x,y$: objects in $X$,
$\forall a,b$: objects in $A$ and $\forall\psi\in X(y,x)$, $\forall f\in
A(a,b)$. Their mutual relations are given by $\eta_{x}=\nu_{x,F(x)}(1_{F(x)}%
)$, $\varepsilon_{a}=\nu_{G(a),a}^{-1}(1_{G(a)})$ $\Longleftrightarrow\nu
(\psi)=\varepsilon F(\psi)$, $\nu^{-1}(f)=G(f)\eta$. (When $\eta$ is
invertible, the adjunction is called an isomorphism and the obstruction for
$\eta$ to be isomorphism yields a cohomology theory.) Identifying
$A=Th/\mathcal{C}(\mathcal{S}_{x})$, $X=E_{x}/\mathcal{S}_{x}$, $x=\omega\in
E_{x}$, $a=\rho_{x}\in Th$, $G=\mathcal{C}^{\ast}$, we apply this to the case
(\ref{adjunction2}) of non-equilibrium local states. Then $F(\omega)$ can be
understood as (the restriction to $\mathcal{S}_{x}$ of) the Hahn-Banach
extension $\nu=\nu_{+}-\nu_{-}\in C(B_{K})^{\ast}$ of $\mathcal{C}%
(\mathcal{S}_{x})\ni\mathcal{C}(\hat{A})\longmapsto\omega(\hat{A})$ to
$\mathcal{T}_{x}$ and $FG=F\mathcal{C}^{\ast}=1_{Th}$. Then $\nu_{-}\neq0$ for
$\omega\notin K$ signals the deviation of $GF=$ $\mathcal{C}^{\ast}F$ from
$1_{E_{x}}$. Therefore, we encounter the \textit{hierarchical family of
adjunctions} according to the choice of $\mathcal{S}_{x}$($\subset
\mathcal{T}_{x}$), in which not only the validity of adjunctions with a
suitable $\mathcal{S}_{x}$ but also its breakdown for a bigger $\mathcal{S}%
_{x}^{\prime}$($\supset\mathcal{S}_{x}$) are physically meaningful.

Next, we recall another important aspect of the adjunction. In decoding the
deep messages encoded in a selection criterion, what plays the decisive roles
at the first stage is the identification of the \textit{centre} of a
representation containing universally all the selected quantum states; its
spectrum provides us with the information on the associated sector structure,
which serves as the vocabulary to be used when the interpretations of a given
quantum state are presented. The necessary bridge between the selected generic
quantum states and the classical familar objects living on the above centre is
provided, in one direction, by the \textit{c}$\rightarrow$\textit{q channel}
which embeds all the known classical states (=probability measures) into the
form of quantum states constituting the totality of the selected states by the
starting selection criterion. The achieved identification between what is
selected and what is embedded from the known world is nothing but the most
important consequence of the categorical adjunction formulated in the form of
selection criterion. This automatically enables us to take the inverse of the
\textit{c}$\rightarrow$\textit{q channel}\ which brings in another most
important ingredient, the \textit{q}$\rightarrow$\textit{c channel} to decode
the physical contents of selected states from the viewpoint of those aspects
selected out by the starting criterion. Mathematically speaking, the spectrum
of the above centre is nothing but the \textit{classifying space }universally
appearing in the geometrical contexts; for instance, in Sec.3 of DR sector
theory of unbroken symmetry described by a compact Lie group $G$, its dual
$\hat{G}$ (of all the equivalence classes of irreducible unitary
representations) is such a case, $\hat{G}=B_{\mathcal{T}}$ for $\mathcal{T}$
the DR category of local endomorphisms of the observable net, where our
\textit{q}$\rightarrow$\textit{c channel }$(\Lambda_{\mu}^{\ast})^{-1}$ plays
the role of the \textit{classifying map} by embedding the $G$-representation
contents of a given quantum state into the subset of $\hat{G}$ consisting of
its irreducible components. For an arbitrary $(\mathfrak{A,}G)$-module
$E=\underset{\gamma\in M}{\oplus}\mathfrak{H}_{\gamma}$ (corresponding to a
choice of state of $\mathfrak{A}$ as in Sec.3) whose $G$-representation
structure is specified by a subset $M$ of $\hat{G}$, we obtain the following
relation in parallel with the definition of classifying maps of $G$-bundles:
\[%
\begin{array}
[c]{ccc}%
\underset{\gamma\in M}{\oplus}\mathfrak{H}_{\gamma}=E &  & \underset{\gamma
\in\hat{G}}{\oplus}\mathfrak{H}_{\gamma}\text{: }%
\begin{array}
[c]{c}%
\text{universal bundle}\\
\text{of all the sectors}%
\end{array}
\\
\text{ \ \ \ \ \ \ \ \ \ \ \ \ \ \ \ \ \ \ \ }\downarrow_{Rep_{G}}\text{ \ } &
& \downarrow_{Rep_{G}}\text{ \ \ \ \ \ \ \ \ \ \ \ \ \ \ \ \ \ \ \ \ }\\
\text{ \ \ \ }(\hat{G}\supset)M & \overset{\text{supp}\circ(\Lambda_{\mu
}^{\ast})^{-1}}{\hookrightarrow} & \hat{G}=B_{\mathcal{T}}\text{ \ :
classifying space}%
\end{array}
.
\]
Corresponding to the relevance of \textit{homotopy} to the situations where
classifying maps appear to reproduce the bundle structure up to homotopy,
everything here is up to multiplicities, since the $G$\textit{-charge
contents} of a selected generic state $\omega$ are examined on the basis of
the data coming from the centre which neglects all the information concerning
the multiplicities. In this way, the present scheme can easily be related with
many current topics concerning the geometric and classification aspects of
commutative as well as non-commutative geometry based upon the (homotopical)
notions of classifying spaces, K-theory and so on.

\subsection{Spectral decomposition and probabilistic interpretation in quantum
measurements}

In view of the importance of the interpretations above, we pick up some
relevant points here from the quantum measurement processes, in regard to the
following basic points:

i) The operator-theoretical notion of spectral decomposition of a self-adjoint
observable $A$ to be measured is equivalent to the algebraic homomorphism
(so-called the map of ``functional calculus''):
\begin{align}
\hat{A} :L^{\infty}(Spec(A))\ni &  f \mapsto\ \hat{A}(f)=f(A)\nonumber\\
&  :={\int}_{a\in Spec(A)} f(a)\ E_{A}(da)\in\mathfrak{A}^{\prime\prime
}\subset B(\mathfrak{H}),
\end{align}
where $\mathfrak{H}$ is the Hilbert space of the defining representation of
the observable algebra $\mathfrak{A}$ to which our observable $A$ belongs.
Here we omit the symbol for discriminating the original C*-algebra
$\mathfrak{A}$ and its representation in $\mathfrak{H}$, and hence, we will
freely move between C*- and W*-versions without explicit mention. This fits
quite well to the common situations of discussing measurements owing to the
absence of disjoint representations in the purely quantum side $\mathfrak{A}$
with \textit{finite} degrees of freedom (due to Stone-von Neumann theorem). In
such cases, the non-trivial existence of a centre comes only from the
classical system coupled to quantum one (, the former of which need to be
derived from the quantum system with infinite degrees of freedom at the
``ultimate'' levels, though).

ii) To give this homomorphism $\hat{A}$ is (almost) equivalent to giving a
spectral measure $E_{A}$ by
\begin{equation}
E_{A}:\mathcal{B}(Spec(A))\ni\Delta\mapsto\ E_{A}(\Delta):=\hat{A}%
(\chi_{\Delta})=\chi_{\Delta}(A)\ \in\ \mathrm{Proj}(\mathfrak{H}),
\end{equation}
on the $\sigma$-algebra $\mathcal{B}(Spec(A))$ on $Spec(A)$ of Borel sets
$\Delta$, identified with the indicator function $\chi_{\Delta}$, taking
values in the set $\mathrm{Proj}(\mathfrak{H})$ of orthogonal projections in
$\mathfrak{H}$. Then the dual map $\hat{A}^{\ast}$ defines a mapping from a
quantum state $\omega$ to a \textit{probability distribution}, $p^{A}%
(\cdot|\ \omega):\mathcal{B}(Spec(A))\ni\ \Delta\ \mapsto\ p^{A}(\Delta
|\omega)=\text{Prob}(A\in\Delta\ |\ \omega):=\omega(E_{A}(\Delta))$, of
measured values in the measurements of $A$ performed in the state $\omega$.
The above reservation ``(almost) equivalent'' is due to the fact that the
reverse direction from a probability distribution to a spectral decomposition
admits a sligthly more general notion, positive-operator valued measure (POM),
which corresponds to a unital completely positive map instead of a
homomorphism and which becomes relevant for treating the set of mutually
non-commutative observables. In any case, the operational meaning of the
mathematical notion of spectral decomposition is exhibited by this $\hat
{A}^{\ast}$ (or, the dual of POM) as a simplest sort of \textit{q}%
$\rightarrow$\textit{c channel} providing the familiar probabilistic interpretation.

iii) To implement physically the spectral decomposition, however, we need some
\textit{physical interaction processes} between the system and the apparatus
through the coupling term of the observable $A\in\mathfrak{A}$ to be measured
and an external field $J$ belonging to the apparatus. While one of the most
polemic issues in the measurement theory is as to how this \textquotedblleft
contraction of wave packets\textquotedblright\ is realized consistently with
the \textquotedblleft standard\textquotedblright\ formulaion of quantum
theory, we here avoid this issue, simply taking such a \textquotedblleft
phenomenological\textquotedblright\ standpoint that our purpose will be
attained if the composite system consisting of the object system and the
classical system involving $J$ is effectively (Fourier- or Legendre-)
transformed through this coupled dynamical process into $\mathfrak{A}\otimes
C^{\ast}\{A\}=:\mathfrak{A}_{A}=C(Spec(A),\mathfrak{A})$, the centre of which
is just the commutative C*-algebra $C^{\ast}\{A\}\simeq C(Spec(A))$ generated
by a self-adjoint operator $A$: $C^{\ast}\{A\}\overset{\iota}{\hookrightarrow
}$ $\mathfrak{Z}(\mathfrak{A}_{A})\hookrightarrow\mathfrak{A}_{A}$. So the
\textit{sector structure} comes in here with sectors parametrized by the
spectrum of the observable $A$ to be measured. (It was the important
contribution of Machida and Namiki \cite{MN} that shed a new light on the
notion of continuous superselection rules, where the focus was, unfortunately,
upon sectors related to \textit{irrelevant unobservable }variables, in sharp
contrast to those discussed here.)

\subsection{Measurement scheme and its realizability}

Then the basic measurement scheme \cite{Ozawa} reduces to the requirement that
all the information on the probability distribution in ii) should be recorded
in and can be read out from this classical part $\{A\}^{\prime\prime
}=L^{\infty}(Spec(A))$ as a mathematical representative of the measuring
apparatus:
\begin{equation}
\omega(E_{A}(\Delta))=p^{A}(\Delta|\omega)=(\omega\otimes\mu_{0})[\hat{\tau
}(\mathbf{1}\otimes\chi_{\Delta})], \label{MeasSchm}%
\end{equation}
where $\mu_{0}$ is some initial state of $\{A\}^{\prime\prime}$ and $\hat
{\tau}\in Aut(\mathfrak{A}_{A})$ describes the effects of dynamics of the
composite system of $\mathfrak{A}$ and $C^{\ast}\{A\}$ (or, more generally, a
dissipative dynamics of a completely positive map also to be allowed).

We are interested here in examining how the problem of a selection criterion
according to our general formulation becomes relevant to the present context.
Applying to any state $\hat{\omega}\in E_{\mathfrak{A}_{A}}$ the uniquely
determined \textit{central decomposition}, we have
\begin{equation}
\hat{\omega}=\int_{Spec(A)}d\mu(a)(\omega_{a}\otimes\delta_{a}),
\label{centrl_dec}%
\end{equation}
with some family of states $\{\omega_{a}\}\subset E_{\mathfrak{A}}$ (which can
be universally chosen by $\omega_{a}(B):=\langle\psi_{a}\ |\ B\psi_{a}\rangle$
with $A\psi_{a}=a\psi_{a}$ \textit{if} $A$ has only discrete spectrum without
multiplicity). What plays important roles here is the instrument
$\mathcal{J}_{A,\tau}$ \cite{Davies-Lewis} depending on $A\in\mathfrak{A}$ and
on the composite-system dynamics $\hat{\tau}$ defined by
\begin{align}
\mathcal{I}_{A,\hat{\tau}}:\mathfrak{A}_{A}\ni\hat{B}\longmapsto
\mathcal{I}_{A,\hat{\tau}}(\hat{B})  &  :=\int d\mu_{0}(a)(\hat{\tau}(\hat
{B}))(a)\nonumber\\
&  =\int d\mu_{0}(a)\delta_{a}(\hat{\tau}(\hat{B}))\in\mathfrak{A,}\\
\mathcal{J}_{A,\hat{\tau}}(\Delta|\omega)(B)  &  :=[\mathcal{I}_{A,\hat{\tau}%
}^{\ast}(\omega)](B\otimes\chi_{\Delta}))=\omega(\mathcal{I}_{A,\hat{\tau}%
}(B\otimes\chi_{\Delta}))\nonumber\\
&  =(\omega\otimes\mu_{0})[\hat{\tau}(B\otimes\chi_{\Delta})].
\end{align}
In terms of these notions, Eq.(\ref{MeasSchm}) can be rewritten as
\begin{align}
\hat{A}^{\ast}(\omega)  &  =(\mathcal{I}_{A,\hat{\tau}}\circ\iota^{\prime
})^{\ast}(\omega)\nonumber\\
\Longrightarrow\hat{A}^{\ast}  &  =\iota^{\prime}{}^{\ast}\circ\mathcal{I}%
_{A,\hat{\tau}}^{\ast}, \label{select}%
\end{align}
where $\iota^{\prime\ast}:E_{\mathfrak{A}_{A}}\rightarrow M_{1}(Spec(A))$
defined by the dual of
\begin{equation}
\iota^{\prime}:\{A\}^{\prime\prime}\ni f\longmapsto\mathbf{1}\otimes
f\in\mathfrak{A}_{A}^{\prime\prime}%
\end{equation}
is the standard (tautological) \textit{q}$\rightarrow$\textit{c channel }to
allow the data read-out from the system-apparatus composite system.
Eq.(\ref{select}) selects out an observable $A$(, or its corresponding
\textit{q}$\rightarrow$\textit{c }channel\textit{\ }$\hat{A}^{\ast}$
describing the probabilistic interpretation of $A$) according to a criterion
as to whether it can be factorized into the standard tautological
\textit{q}$\rightarrow$\textit{c channel }$\iota^{\prime\ast}$ and some
instrument $\mathcal{I}_{A,\hat{\tau}}$. In view of the formal similarity
between the relation $\pi_{\omega}=\pi_{0}\circ\rho$ coming from the DHR
criterion and Eq.(\ref{select}), it is interesting to note that what are
examined here is \textit{q}$\rightarrow$\textit{c }channels, $\hat{A}^{\ast}$
and $\iota^{\prime}{}^{\ast}$, the latter of which is a fixed standard one.
This criterion is just for examining whether the measurement of $A$ can
actually be materialized by means of the coupling $\hat{\tau}$ between the
system containing $A$ and some measuring apparatus constituting the composite
system $\mathfrak{A}_{A}=\mathfrak{A}\otimes\{A\}^{\prime\prime}$. In this
sense, the criterion examines the \textit{realization problem }in the context
of control theory \cite{ArbibMane}, asking whether a suitable choice of an
apparatus and a choice of dynamical coupling can correctly describe the
input-output behaviour of the system.\textit{\ }Once this criterion is valid,
its experimental observation is most conveniently described by the instrument
$\mathcal{J}_{A,\hat{\tau}}(\Delta|\omega)(B)$ whose interpretation is given
\cite{Ozawa} by

1) the probability distribution of the measured value of $A$ in a state
$\omega$ is given by $\mathcal{J}_{A,\hat{\tau}}(\Delta|\omega)(\mathbf{1}%
)=p_{A}(\Delta|\omega),$

2) the final state realized (in the repeatable measurement) after the readout
$a\in\Delta$ is given by the Radon-Nikodym derivative $\mathcal{J}%
_{A,\hat{\tau}}(da|\omega)/p_{A}(da|\omega)$,

3) in combination of 1) and 2), the quantity $\mathcal{J}_{A,\hat{\tau}%
}(\Delta|\omega)(B)$ itself can be regarded as the expectation value of
another observable $B\in\mathfrak{A}$ when the initial state $\omega$ goes
into some final state whose $A$-values belong to $\Delta(\subset Spec(A))$.

\subsection{Problem of state preparation as reachability problem}

In the related context, we need to examine the problem of
\textit{reachability} to ask whether there is a controlled way to drive the
(composite) system to any desired state starting from some initial state; this
is nothing but the problem of \textbf{state preparation}, which has not been
seriously discussed, in spite of its vital importance in the physical
interpretation of quantum theory.

For this purpose, we need to define the \textit{c}$\rightarrow$\textit{q
channel} relevant to it. Fixing a family $(\omega_{a})_{a\in Spec(A)}=:\phi$
of states on $\mathfrak{A}$ appearing in the central decomposition
(\ref{centrl_dec}), we can define a \textit{c}$\rightarrow$\textit{q channel}
by
\begin{equation}
C_{A,\phi}:\mathfrak{A}_{A}\ni\hat{B}\longmapsto(Spec(A)\ni a\longmapsto
\omega_{a}(\hat{B}(a)))\in C(Spec(A)),
\end{equation}
and hence, $C_{A,\phi}^{\ast}:M_{1}(Spec(A))\ni\rho\longmapsto C_{A,\phi
}^{\ast}(\rho)\in E_{\mathfrak{A}_{A}}$, where
\begin{align}
C_{A,\phi}^{\ast}(\rho)(\hat{B})  &  =\rho(C_{A,\phi}(\hat{B}))=\int
d\rho(a)\omega_{a}(\hat{B}(a))=\int d\rho(a)(\omega_{a}\otimes\delta_{a}%
)(\hat{B}),\nonumber\\
\text{or, }C_{A,\phi}^{\ast}(\rho)  &  =\int d\rho(a)(\omega_{a}\otimes
\delta_{a}).
\end{align}
In terms of these, the reachability (or, preparability) criterion can be
formulated as the problem to examine the validity of
\begin{equation}
\omega=\lim_{t\rightarrow\infty}(\iota^{\ast}\circ C_{A,\phi}^{\ast}%
)(\mu_{\hat{\tau}_{t}}), \label{prepare}%
\end{equation}
where $\iota^{\ast}:E_{\mathfrak{A}_{A}}\rightarrow E_{\mathfrak{A}}$ is the
dual of $\iota:\mathfrak{A}\ni B\longmapsto B\otimes\mathbf{1}\in
\mathfrak{A}_{A}$, and the measure $\mu_{\hat{\tau}_{t}}^{\omega}\in
M_{1}(Spec(A))$ is defined through the central decomposition of $(\omega
\otimes\mu_{0})\circ\hat{\tau}_{t}=\int d\mu_{\hat{\tau}_{t}}^{\omega
}(a)\omega_{a}\otimes\delta_{a}$ valid for such an observable $A$ as with
discrete spectrum. If we can find such a suitable coupled dynamics $\hat{\tau
}_{t}$ and an initial and final probability measures $\mu_{0},\mu_{1}\in$
$M_{1}(Spec(A))$ that $\lim_{t\rightarrow\infty}(\omega\otimes\mu_{0}%
)\circ\hat{\tau}_{t}(B\otimes\mathbf{1})=(\omega\otimes\mu_{1})(B\otimes
\mathbf{1})$ for each $B\in\mathfrak{A}$, then a state $\omega$ can actually
be prepared:
\begin{align}
&  (\iota^{\ast}\circ C_{A,\phi}^{\ast})(\mu_{\hat{\tau}})(B)=\mu_{\hat{\tau
}_{t}}(C_{A,\phi}(B\otimes\mathbf{1}))=\int d\mu_{\hat{\tau}_{t}}(a)\omega
_{a}(B\otimes\mathbf{1})\nonumber\\
&  =(\omega\otimes\mu_{0})\circ\hat{\tau}_{t}(B\otimes\mathbf{1}%
)\underset{t\rightarrow\infty}{\rightarrow}(\omega\otimes\mu_{1}%
)(B\otimes\mathbf{1})=\omega(B),
\end{align}
in the sense that there is some operational means specified in terms of
$A\in\mathfrak{A}$, a coupled dynamics $\hat{\tau}_{t}$ and an initial and
final probability measures $\mu_{0},\mu_{1}\in$ $M_{1}(Spec(A))$.

Here, the assumption of discreteness of the spectrum of $A$ is no problem,
since $A$ plays here only a subsidiary role. However, this problem becomes
crucial when we start to examine the \textit{repeatability} of the measurement
of the observable $A$ itself. We compare the above \textit{q}$\rightarrow
$\textit{c channel} $(C_{A,\phi}^{\ast})^{-1}$ with another natural
\textit{q}$\rightarrow$\textit{c channel} $(\iota\circ\hat{A})^{\ast}$, which
can be defined on all the states $\in E_{\mathfrak{A}_{A}}$, independently of
a specific choice of a family $\phi=(\omega_{a})_{a\in Spec(A)}$ of states on
$\mathfrak{A}$, simply as the dual of the composed embedding maps,
$C(Spec(A))\overset{\hat{A}}{\hookrightarrow}\mathfrak{A}\overset{\iota
}{\hookrightarrow}\mathfrak{A}_{A}$. As is seen from the relation,
\begin{align}
&  (\iota\circ\hat{A})^{\ast}(\int d\mu(a)(\omega_{a}\text{ }\otimes\delta
_{a}))(f)\nonumber\\
&  =\int d\mu(a)(\omega_{a}\text{ }\otimes\delta_{a}))((\iota\circ\hat
{A})(f))=\int d\mu(a)(\omega_{a}\text{ }\otimes\delta_{a}))(f(A)\otimes
\mathbf{1})\nonumber\\
&  =\int d\mu(a)\omega_{a}(f(A))=\int d\mu(a)\int\omega_{a}(dE_{A}(b))f(b),
\end{align}
$(\iota\circ\hat{A})^{\ast}$ is, in general, not equal to $(C_{A,\phi}^{\ast
})^{-1}$, nor has a simple interpretation. If we can choose such a family
$(\omega_{a})_{a\in Spec(A)}$ that $\int f(b)\omega_{a}(dE_{A}(b))=f(a)$ for
$\forall f\in C(Spec(A))$, or equivalently, $\omega_{a}(E_{A}(\Delta
))=\chi_{\Delta}(a)$ for $\forall\Delta$: measurable subset of $Spec(A)$, we
can attain the equality between $(C_{A,\phi}^{\ast})^{-1}$ and $(\iota
\circ\hat{A})^{\ast}$ on the image of $C_{A,\phi}^{\ast}$ in $E_{\mathfrak{A}%
_{A}}$, which can be extended to the whole $E_{\mathfrak{A}_{A}}$ by the use
of the Hahn-Banach extension. As a result, we can attain universally the state
preparations and physical interpretations (in relation to $A$), independently
of a specific choice of the above family $(\omega_{a})_{a\in Spec(A)}$. While
such a choice is always possible for observables $A$ with discrete spectrum,
its impossibility for those $A$ with \textit{continuous spectra} forces us to
consider the \textit{approximate measurement scheme} (see \cite{Ozawa}), which
involves the essential dependence on the choice of the family $(\omega
_{a})_{a\in Spec(A)}$ and the selection of and restriction to preparable and
interpretable states.

In this way, we have seen that this approach provides a simple unified scheme
based upon instruments and channels for discussing various aspects in the
measurement processes without being trapped in the depth of philosophical
issues. So, it will be worthwhile to attempt the possible extension of the
measurement scheme to more general situations involving QFT. It will be also
interesting to examine the problems of state correlations in entanglements, of
state estimation, and so on, in use of the notions of mutual entropy, channel
capacities \cite{Ohya, OhyaPetz}, Cram\'{e}r-Rao bounds, etc.

Through the above relation with the spectral decomposition of an observable
$A$ and the superselection sectors parametrized by $a\in Spec(A)$, we can
reconfirm the naturalility of our extending the meaning of sectors from their
traditional version of discrete one, to the present version including both: in
SSB, order parameters of continuous family of disjoint states (of
$\mathfrak{A}$) parametrized by $H\backslash G$ and in thermal situations,
(inverse) temperatures $\beta$[=$(\beta^{\mu})$] discriminating pure
thermodynamic phases corresponding also to disjoint KMS states (of
$\mathfrak{A}$), and variety of non-equilibrium local states (\cite{BOR01}).
Our way of unifying these various cases is seen to be quite similar to the
unified treatment of discrete and continuous spectra of self-adjoint operators
in the general theory of spectral decompositions.

We conclude this paper by mentioning some problems under investigation, which
will be reported somewhere.

\begin{itemize}
\item[1.] Treatment of a non-compact group of broken internal symmetry as
remarked in Sec.4.3 and 4.4.

\item[2.] Reformulation of characterization of KMS states: in Sec.2, we have
just relied on the known simplicial structure of the set of all KMS states. To
be consistent with the spirit of the present scheme, we need also to find a
version of selection criterion to characterize these KMS states, whose essence
should be found in the \textbf{zeroth law of thermodynamics }from which the
familiar parameter of temperature arises (in combination with the first and
second laws in such a form as the passivity \cite{PuszWoro} or the Gibbs
variational principle \cite{ArakiMoriya}). In any case, such a physically
interesting problem as drawing a phase diagram to accommodate phase
transitions just belongs to the analysis of sector structure in the present context.

\item[3.] To substantiate the above consideration, it is necessary to develop
a systematic way of treating a chemical potential \cite{AHKT} as one of the
order parameters to be added to temperature. This requires the local and
systematic treatment of conserved currents such as $T_{\mu\nu}$(:
energy-momentum tensor) and $j_{\mu}$(: current density), extended to thermal
situations just in a parallel way to the local thermal observables in
\cite{BOR01}. To understand sectors in relation with spacetime structure, the
notion of soliton sectors \cite{Froeh76} seems also quite interesting.

\item[4.] It would be worthwhile to examine whether the notion of a field
algebra $\mathfrak{F}$ is a simple mathematical device, convenient for making
the interpretation easier from the viewpoint laid out by Klein's Erlangen
programme and no more than that.
\end{itemize}

\section*{Acknowledgments}

I am very grateful to Prof.~S.~Maumary and Institut de math\'{e}matiques,
Universit\'{e} de Lausanne, for their kind hospitality and supports during my
stays in 2001 and 2002, where some part of this work was done. I would like to
express my sincere thanks to Prof. J.~Fr\"{o}hlich for his careful reading of
the manuscript as well as constructive comments to an earlier version of this
paper. I am very grateful also to Profs.~R.~Haag and D.~Sternheimer for
valuable discussions and their unchanged warm encouragements in my current
projects. I deeply thank Prof.~J.~E.~Roberts for his critical comments
pointing out some errors in the early versions.

\end{document}